\documentclass[prd,aps,nofootinbib,notitlepage,showpacs,preprintnumbers]{revtex4-1}
\usepackage{graphicx,epsf,amsmath,amsfonts,amssymb,amsbsy}
\usepackage{epsfig}
\newcommand{\ds}{\displaystyle}
\newcommand{\vev}[1]{\langle#1\rangle}
\newcommand{\mat}{\left ( \begin{array}}
\newcommand{\emat}{\end{array} \right )}
\newcommand{\vect}{\left ( \begin{array}{c}}
\newcommand{\evect}{\end{array} \right )}

\begin{document}

%\hfill HU-EP-11/11

\title{Magnetic catalysis effect in the (2+1)-dimensional Gross--Neveu model with Zeeman interaction }
\author{K.G. Klimenko $^{a,b}$,
R.N. Zhokhov $^{a}$ }%and V.Ch. Zhukovsky $^{c}$}
\affiliation{$^{a}$ Institute for High Energy Physics, 142281,
Protvino, Moscow Region, Russia} \affiliation{$^{b}$ University
"Dubna" (Protvino branch), 142281, Protvino, Moscow Region, Russia}
%\affiliation{$^{c}$ Faculty of
%Physics, Moscow State University, 119991, Moscow, Russia}

\begin{abstract}
Magnetic catalysis of the chiral symmetry breaking and other
magnetic properties of the (2+1)-dimensional Gross--Neveu model are
studied taking into account the Zeeman interaction of spin-1/2
quasi-particles (electrons) with tilted (with respect to a system
plane) external magnetic field $\vec B=\vec B_\perp+\vec
B_\parallel$. The Zeeman interaction is proportional to magnetic
moment $\mu_B$ of electrons. For simplicity,
temperature and chemical potential are equal to zero throughout the
paper. We compare in the framework of the model the above mentioned
phenomena both at $\mu_B=0$ and $\mu_B\ne 0$. It is shown that at
$\mu_B\ne 0$ the magnetic catalysis effect is drastically changed in
comparison with the $\mu_B= 0$ case. Namely, at $\mu_B\ne 0$ the
chiral symmetry, being spontaneously broken by $\vec B$ at
subcritical coupling constants,
%or enhanced at rather small values of $|\vec B|$,
is always restored at $|\vec B|\to\infty$ (even at $\vec
B_\parallel=0$). Moreover, it is proved in this case that chiral
symmetry can be restored simply by tilting $\vec B$ to a system
plane, and in the region $ B_\perp\to 0$ the de Haas -- van Alphen
oscillations of the magnetization are observed. At supercritical
values of coupling constant we have found two chirally non-invariant
phases which respond differently on the action of $\vec B$. The
first (at rather small values of $|\vec B|$) is a diamagnetic phase,
in which there is an enhancement of chiral condensate, whereas the
second  is a paramagnetic chirally broken phase. Numerical estimates
show that phase transitions described in the paper can be achieved
at low enough laboratory magnetic fields.
\end{abstract}

%%% 12.38.Mh Quark-gluon plasma
%%% 21.65.Qr Quark matter
%%% 12.39.Ki Relativistic quark model
%\pacs{12.39.Ki, 12.38.Mh, 21.65.Qr}

%%% 12.38.Mh Quark-gluon plasma
%%% 21.65.Qr Quark matter
%%% 12.39.Ki Relativistic quark model

\maketitle

\section{ Introduction}

It is well known that during last three decades a lot of attention
is paid to the investigation of (2+1)-dimensional quantum field
theories (QFT) under influence of different external conditions. In
particular, the (2+1)-dimensional Gross-Neveu (GN) \cite{gn} type
models are among the most popular
\cite{semenoff,rosenstein,klimenko3}.
%Last time a lot of attention is paid to the consideration of QFT in two spatial dimensions.
There are several basic motivations for this interest. Since low
dimensional theories have a rather simple structure, they can be
used in order to develop our physical intuition for different
physical phenomena taking place in real (3+1)-dimensional world
(such as dynamical symmetry breaking
\cite{gn,semenoff,rosenstein,klimenko3,
inagaki,hands,bashir,khanna,gies}, color superconductivity
\cite{toki} etc). Another example of this kind is the spontaneous
chiral symmetry breaking induced by external magnetic fields, i.e.
the magnetic catalysis effect (see the recent reviews
\cite{shovkovy,incera} and references therein). For the first time
this effect was also studied in terms of (2+1)-dimensional GN models
\cite{klimenko}. In addition, low dimensional models are useful in
elaborating new QFT methods like the large-$N$ technique
\cite{gn,rosenstein} and the optimized expansion method
\cite{kneur,k} etc.

However, a more fundamental reason for the study of these theories
is also well known. Indeed, there are a lot of condensed matter
systems which, firstly, have a (quasi-)planar structure and,
secondly, their low-energy excitation spectrum is described
adequately by relativistic Dirac-like equation rather than by
Schr\"{o}dinger one. Among these systems are the high-T$_c$ cuprate
and iron superconductors \cite{davydov}, the one-atom thick layer of
carbon atoms, or graphene, \cite{niemi,castroneto} etc. Thus, many
properties of such condensed matter systems can be explained in the
framework of various (2+1)-dimensional QFTs, including the GN-type
models (see, e.g.,
\cite{Semenoff:1998bk,babaev,Fialkovsky,caldas,zkke,marino,zhokhov,gies2,herbut}
and references therein). Especially, it is necessary to note that
the magnetic catalysis phenomenon supplies a very effective
mechanism for a description of high temperature superconductivity
\cite{Semenoff:1998bk,zkke}, quantum Hall effect in graphene
\cite{gorbar} and other condensed matter systems, which have a thin
film structure and are exposed to an external magnetic field.

Since elementary excitations of an arbitrary condensed matter system
are usually electrons (or quasi-particles) with $\pm 1/2$ spin
projection on the direction of external magnetic field $\vec B$,
there are two independent ways to introduce $\vec B$ into
consideration in such systems, i) when $\vec B$ couples only to an
orbital angular momentum of electrons (note, in planar systems only
perpendicular component $\vec B_\perp$ of external magnetic
field $\vec B$ contributes in this case), and ii) when, in addition,
the Zeeman interaction of an electron spin (or its magnetic moment)
with $\vec B$ is also taken into account. In early investigations of
the magnetic catalysis effect in the framework of (2+1)-dimensional
GN type models (see, e.g., \cite{klimenko,Semenoff:1998bk,zkke})
just the first way i) was used, where it was shown that chiral
symmetry of the model is spontaneously broken down at arbitrary
values of $\vec B$ and even at infinitesimal values of the coupling constant. Later, especially in connection with graphene physics, the
Zeeman interaction was also taken into account (see, e.g.,
\cite{gorbar,Slizovskiy:2012kd}), but only in the case when $\vec B$
is perpendicular to the system plane. In contrast, in the recent
paper \cite{caldas} another limiting case for the direction of an
external magnetic field was considered. There the influence of
in-plane magnetic field $\vec B_\parallel$, i.e. parallel to the
system plane, on the (2+1)-dimensional GN model was studied (in this
case, since $\vec B_\perp=0$, the magnetic field couples to fermions
only due to the Zeeman interaction). In particular, it was shown in
\cite{caldas} that at sufficiently high values of $\vec B_\parallel$
chiral symmetry of the model is restored. Naturally, one might
wonder about the response of the chiral symmetry of
(2+1)-dimensional GN type models upon the action of an external
arbitrarily {\it directed} magnetic field, when $\vec B$ interacts
both with orbital angular momentum and spin of electrons. (In this
connection it is worth to note that in the framework of a planar
system of free electrons a jumping behavior of the magnetization vs
$|\vec B|$ was predicted in \cite{schakel} at rather small
(laboratory) values of $|\vec B|$. The effect was explained in
\cite{schakel} just due to the Zeeman interaction of spin with {\it
tilted} magnetic field.)

So, in the present paper the magnetic catalysis effect, as well as
other magnetic phenomena, is investigated in the leading order of
the large-$N$ expansion technique in the framework of the
(2+1)-dimensional GN model subjected to a tilted external magnetic
field $\vec B$. The model is invariant with respect to the discrete
chiral symmetry and describes the four-fermion interaction of
quasi-particles (electrons) with ($\pm 1/2$)-spin projections on the
direction of $\vec B$. The Zeeman interaction of an electron
magnetic moment with $\vec B$ is also taken into account.
Temperature and chemical potential are equal to zero throughout the
consideration. In particular, we show that at subcritical values of
the coupling constant the chiral symmetry, being broken
spontaneously at rather small values of $\vec B$, is restored (in
contrast to the case when Zeeman interaction is ignored
\cite{klimenko,Semenoff:1998bk,zkke}) at sufficiently high values of
external magnetic field. Moreover, we have found a jumping
behavior of dynamical fermion mass vs $|\vec B|$, i.e. the evolution
of the system vs $\vec B$ is accompanied by its passing through
several different phases with broken chiral symmetry. It turns out
that due to a presence of the Zeeman interaction one can observe in
the model the de Haas -- van Alphen  oscillations of the
magnetization as well as diamagnetic and paramagnetic phenomena. We
hope that our results can be useful in explaining physical phenomena
in thin organic films, graphene etc, i.e. in all condensed matter
systems having a spatially (quasi-) two-dimensional structure.

The paper is organized as follows. In Sec. II the (2+1)-dimensional
GN model is presented as well as its renormalized thermodynamic
potential is obtained in different particular cases, i.e. with- and
without Zeeman interaction, when external magnetic field is taken
into account. In the next Sec. III the modification of the magnetic
catalysis effect is investigated in the case of subcritical values
of the coupling constant (Sec. III A). Here it is also shown (Sec.
III B) that tilting of the magnetic field leads to 
oscillations of the magnetization. In Sec. IV A and IV B the case of
supercritical values of the coupling constant is analyzed. It is
established here that the phase portrait of the model contains two
chirally non-invariant phases. One of them has a diamagnetic ground
state, in another phase it is a paramagnetic one. It is also shown
that at sufficiently high values of $|\vec B|$ and at $B_\perp\to 0$
chiral symmetry of the model is restored. In Sec. IV C some
numerical estimates in the context of condensed matter systems are
performed, which show that phase transitions induced by Zeeman
effect are really achieved at laboratory magnitudes of external
magnetic fields. Finally, in Sec. V the summary of the paper is
presented.

\section{ The model and its thermodynamic potential}
\label{effaction}

We suppose that some physical system is localized in the spatially two-dimensional plane perpendicular to the $\hat z$ coordinate axis of usual tree-dimensional space. Moreover, there is an
external homogeneous and time independent magnetic field $\vec B$
tilted with respect to this plane. The corresponding
(3+1)-dimensional vector potential $A_\mu$ is given by $A_{0,1}=0$,
$A_2=B_\perp x$, $A_3=B_\parallel y$, i.e. the spatial components
$B_{x,y,z}$ of an external magnetic field have the form
$B_x=B_\parallel$, $B_y=0$, $B_z=B_\perp$. We assume that the planar
physical system consists of quasi-particles (electrons) with two
spin projections, $\pm$1/2, on the direction of magnetic field
$\vec B$. Moreover, it is also supposed that their low-energy dynamics is described by the following (2+1)-dimensional Gross-Neveu type
Lagrangian \footnote{In the paper we use the natural units,
$\hbar=k_B=c=1$.  However, for a more adequate application of our
results to condensed matter physics it is necessary to modify
slightly Lagrangian (1). Namely, one should use there the
replacements $\nabla_{1,2}\to v_F\nabla_{1,2}$ as well as $G\to
v_FG$, where $v_F$ is a Fermi velocity of quasi-particles (for
example, in graphene $v_F\approx c/300$). For simplicity, in all the
expressions of our paper we assume that $v_F=1$. The case $v_F\ne 1$
is considered in the last section IV C, where some estimates are
made in the context of condensed matter systems.}
\begin{eqnarray}
 L=\sum_{k=1}^2\bar \psi_{ka}\Big [\gamma^0i\partial_t+\gamma^1 i\nabla_1+\gamma^2 i\nabla_2
 -\nu(-1)^k\gamma^0\Big ]\psi_{ka}+ \frac{G}{N}\left
(\sum_{k=1}^{2}\bar \psi_{ka}\psi_{ka}\right )^2, \label{1}
\end{eqnarray}
where $\nabla_{1,2}=\partial_{1,2}+ieA_{1,2}$ and the summation over
the repeated index $a=1,...,N$ of the internal $O(N)$ group is
implied. For each fixed value of $k=1,2$ and $a=1,...,N$ the
quantity $\psi_{ka}(x)$ in (1) means the Dirac fermion field,
transforming over a reducible 4-component spinor representation of
the (2+1)-dimensional Lorentz group. Moreover, all these Dirac
fields $\psi_{ka}(x)$ form two fundamental multiplets,
$\psi_{1a}(x)$ and $\psi_{2a}(x)$ ($a=1,...,N$), of the internal
auxiliary $O(N)$ group, which is introduced here in order to make it
possible to perform all the calculations in the framework of the
nonperturbative large-$N$ expansion method. We suppose that spinor
fields $\psi_{1a}(x)$ and $\psi_{2a}(x)$ ($a=1,...,N$) correspond to
electrons with spin projections 1/2 and -1/2 on the direction of an
external magnetic field, respectively. In (1) the $\nu$-term is
introduced in order to take into account the Zeeman interaction
energy of electrons with external magnetic field $\vec B$. (To
investigate the genuine role of a magnetic field in the phase
structure of the model, we suppose throughout the paper that both
electron number chemical potential and temperature are zero.) Hence,
in our case $\nu=g_S\mu_B |\vec B|/2$, where $|\vec
B|=\sqrt{B^2_\parallel+B^2_\perp}$, $g_S$ is the spectroscopic Lande
factor and $\mu_B$ is an electron magnetic moment, i.e. the Bohr magneton.
%%%%%%%
\footnote{In our consideration the Zeeman $\nu$-term in (1) is
introduced phenomenologically, so the quantity $\mu_B$ might be
considered as a free model parameter. However, at the end of the
paper, when doing some numerical estimates in Sec. IV C, we suppose that $\mu_B$ is equal to the Bohr magneton. Recall, there is an interesting possibility of
a dynamical generation of an electron magnetic moment \cite{ferrer}.
This approach, however, is outside of the scope of the present
paper. }
%%%%%%%%%
In what follows  it is supposed  that $g_S=2$, however at the end of
the paper we will also briefly discuss the influence of $g_S>2$, as
in \cite{schakel}, on some physical results. The algebra of the
$\gamma^\rho$-matrices as well as their particular representations
are given, e.g., in \cite{zhokhov}. The model (1) is invariant under
the discrete chiral transformation, $\psi_{ka}\to\gamma^5\psi_{ka}$
(the particular realization of the $\gamma^5$-matrix is also
presented in \cite{zhokhov}). Certainly, there is the $O(N)$
invariance of the Lagrangian (1). Finally note that at $N=1$ the
quasi-particle spectrum of the model (1) is just the same as in the
monolayer graphene \cite{gorbar}, but at $N>1$ one can interpret our
results as occurring in the $N$-layered system.

In the following we use an auxiliary theory with the Lagrangian density
\begin{eqnarray}
{\cal L}\ds =
 -\frac{N\sigma^2}{4G}+\sum_{k=1}^2\bar\psi_{ka}\Big (\gamma^0i\partial_t+
 \gamma^1 i\nabla_1+\gamma^2 i\nabla_2
 +\mu_k\gamma^0 -\sigma \Big )\psi_{ka}, \label{2}
\end{eqnarray}
where $\mu_1=\nu$, $\mu_2=-\nu$ and from now on $\nu=\mu_B |\vec B|$
(in this formula and below the summation over repeated indices is
implied). Clearly, the Lagrangians (\ref{1}) and (\ref{2}) are
equivalent, as can be seen by using the Euler-Lagrange equation of
motion for scalar bosonic field $\sigma (x)$ which takes the form
\begin{eqnarray}
\sigma (x)=-\frac{2G}{N}\sum_{k=1}^2\bar\psi_{ka}\psi_{ka}.
\label{3}
\end{eqnarray}
One can easily see from (\ref{3}) that the neutral field $\sigma(x)$
is a real quantity, i.e. $(\sigma(x))^\dagger=\sigma(x)$ (the
superscript symbol $\dagger$ denotes the Hermitian conjugation).
Moreover, if $\vev{\sigma (x)}\ne 0$, then, after the shifting
$\sigma (x)\to\sigma (x)+\vev{\sigma (x)}$ in (\ref{2}), it is clear
that the discrete chiral symmetry of the model is spontaneously
broken and fermions acquire dynamically the mass equal to
$\vev{\sigma (x)}$.

Let us now study the phase structure of the four-fermion model (1)
by starting from the equivalent semi-bosonized Lagrangian (\ref{2}).
In the leading order of the large-$N$ approximation, the effective
action ${\cal S}_{\rm {eff}}(\sigma)$ of the
considered model is expressed by means of the path integral over
fermion fields
$$
\exp(i {\cal S}_{\rm {eff}}(\sigma))=
  \int\prod_{k=1}^{2}\prod_{a=1}^{N}[d\bar\psi_{ka}][d\psi_{ka}]\exp\Bigl(i\int {\cal
  L}\,d^3 x\Bigr),
$$
where
\begin{eqnarray}
&&{\cal S}_{\rm {eff}} (\sigma) =-\int d^3x\frac{N}{4G}\sigma^2(x)+
\widetilde {\cal S}_{\rm {eff}}. \label{5}
\end{eqnarray}
The fermion contribution to the effective action, i.e.\  the term
$\widetilde {\cal S}_{\rm {eff}}$ in (\ref{5}), is given by
\begin{eqnarray}
\exp(i\widetilde {\cal S}_{\rm
{eff}})&=&\int\prod_{l=1}^{2}\prod_{a=1}^{N}[d\bar\psi_{la}][d\psi_{la}]\exp\Bigl\{
i\int\sum_{k=1}^2\bar\psi_{ka}\Big (\gamma^0i\partial_t+\gamma^1
i\nabla_1+\gamma^2 i\nabla_2 +\mu_k\gamma^0 -\sigma \Big )\psi_{ka}
d^3 x\Bigr\}. \label{6}
\end{eqnarray}
The ground state expectation value $\vev{\sigma(x)}$ of the
composite bosonic field is determined by the saddle point equation,
\begin{eqnarray}
\frac{\delta {\cal S}_{\rm {eff}}}{\delta\sigma (x)}=0. \label{7}
\end{eqnarray}
For simplicity, throughout the paper we suppose that the above
mentioned ground state expectation value does not depend on space-time
coordinates, i.e.
\begin{eqnarray}
\vev{\sigma(x)}\equiv M, \label{8}
\end{eqnarray}
where $M$ is a constant quantity. In fact, it is a
coordinate of the global minimum point of the thermodynamic
potential (TDP) $\Omega (M;\nu,B_\perp)$. In the leading order
of the large-$N$ expansion the TDP is defined by the following
expression:
\begin{equation*}
\int d^3x \Omega (M;\nu,B_\perp)=-\frac 1N{\cal S}_{\rm
{eff}}(\sigma(x))\Big|_{\sigma
    (x)=M} ,
\end{equation*}
which gives
\begin{eqnarray}
\int d^3x\Omega (M;\nu,B_\perp)\,\,&=&\,\,\int
d^3x\frac{M^2}{4G}+\frac iN\ln\left (
\int\prod_{l=1}^{2}\prod_{b=1}^{N}[d\bar\psi_{lb}][d\psi_{lb}]\exp\Big
(i\int\sum_{k=1}^{2}\bar\psi_{ka} D_k\psi_{ka} d^3 x \Big )\right ),
\label{9}
\end{eqnarray}
where $D_k=\gamma^0i\partial_t+\gamma^1 i\nabla_1+\gamma^2 i\nabla_2
+\mu_k\gamma^0-M$.

\subsection{The particular case $B_\perp=0$, $\nu\ne 0$}

In order to find a convenient expression for the TDP in this case
it is necessary to evaluate the Gaussian path integral (\ref{9}) at
$B_\perp=0$ (see, e.g., the paper \cite{zhokhov}, where more general
path integrals with difermion condensates were calculated). As a
result, we obtain the following expression for the  TDP of the model
(1) at zero temperature:
\begin{eqnarray}
\Omega (M;\nu)= \frac{M^2}{4G} +2i\int\frac{d^3p}{(2\pi)^3}\ln\Big
[(p_0^2-(E+\nu)^2)(p_0^2 -(E-\nu)^2)\Big ], \label{12}
\end{eqnarray}
where $E=\sqrt{M^2+|\vec p|^2}$ and $|\vec p|=\sqrt{p_1^2+p_2^2}$.
It is clear from (\ref{12}) that without loss of generality one can
suppose that $\nu\ge 0$ and $M\ge 0$. Using in the expression
(\ref{12}) a rather general formula
\begin{eqnarray}
\int_{-\infty}^\infty dp_0\ln\big
(p_0-A)=\mathrm{i}\pi|A|\label{int}
\end{eqnarray}
(obtained rigorously, e.g., in Appendix B of \cite{gubina} and true
up to an infinite term independent on the real quantity $A$), it is
possible to reduce it to the following one:
\begin{eqnarray}
\Omega (M;\nu)\equiv\Omega^{un} (M;\nu)=
\frac{M^2}{4G}-2\int\frac{d^2p}{(2\pi)^2}\Big (|E+\nu|+|E-\nu|\Big
).  \label{13}
\end{eqnarray}
The integral term in (\ref{13}) is an ultraviolet divergent one, hence
to obtain any information from this expression we have to
renormalize it. First of all, let us regularize the TDP (\ref{13})
by cutting the integration region, i.e. we suppose that $|p_1|<\Lambda$,
$|p_2|<\Lambda$ in (\ref{13}). As a result we have the following
regularized expression (which is finite at finite values of
$\Lambda$):
\begin{eqnarray}
\Omega^{reg} (M;\nu)= \frac{M^2}{4G} -\frac{2}{\pi^2}\int_0^\Lambda
dp_1\int_0^\Lambda dp_2\Big (|E+\nu|+|E-\nu|\Big ). \label{15}
\end{eqnarray}
Let us use in (\ref{15}) the following asymptotic expansion at
$|\vec p|\to\infty$, i.e. at $|\vec p|\gg\nu$,
\begin{eqnarray}
|E+\nu|+|E-\nu|=2|\vec p| +\frac{M^2}{|\vec p|}+{\cal O}(1/|\vec
p|^3).\label{16}
\end{eqnarray}
(Note, the leading asymptotic terms in (\ref{16}) do not depend on
$\nu$.) Then, upon integration there term-by-term, it is possible to
find
 \begin{eqnarray}
\Omega^{reg}(M;\nu)&=&M^2\left [\frac
1{4G}-\frac{4\Lambda\ln(1+\sqrt{2})}{\pi^2}\right
]-\frac{4\Lambda^3(\sqrt{2}+\ln(1+\sqrt{2}))}{3\pi^2}+{\cal
O}(\Lambda^0), \label{17}
\end{eqnarray}
where ${\cal O}(\Lambda^0)$ denotes an expression which is finite in
the limit $\Lambda\to \infty$.  Second, we suppose that the bare
coupling constant $G$ depends on the cutoff parameter $\Lambda$ in
such a way that in the limit $\Lambda\to\infty$ one obtains a finite
expression in the square brackets of (\ref{17}). Clearly, to fulfil
this requirement it is sufficient to require that
 \begin{eqnarray}
\frac 1{4G}\equiv \frac
1{4G(\Lambda)}=\frac{4\Lambda\ln(1+\sqrt{2})}{\pi^2}+\frac{1}{\pi
g}\equiv\frac{1}{4G_c}+\frac{1}{\pi g}, \label{18}
\end{eqnarray}
where $g$ is a finite and $\Lambda$-independent model parameter with
dimensionality of inverse mass and $G_c=\frac{\pi^2}{16\Lambda\ln
(1+\sqrt{2})}$. Moreover, since bare coupling $G$ does not depend on
a normalization point, the same property is also valid for $g$.
Hence, taking into account in (\ref{15}) and (\ref{17}) the relation
(\ref{18}) and ignoring there an infinite $M$-independent constant,
one obtains the following {\it renormalized}, i.e. finite,
expression for the TDP
\begin{eqnarray}
\Omega^{ren}(M;\nu)&=&\lim_{\Lambda\to\infty}
\left\{\Omega^{reg}(M;\nu)\Big |_{G=
G(\Lambda)}+\frac{4\Lambda^3(\sqrt{2}+\ln(1+\sqrt{2}))}{3\pi^2}\right\}.\label{19}
\end{eqnarray}
It should also be mentioned that the TDP (\ref{19}) is a
renormalization group invariant quantity.

Suppose that $\nu\equiv \mu_B |\vec B|=0$. Then the ${\cal
O}(\Lambda^0)$ term in (\ref{17}) can be calculated explicitly. As a
result, we have for the TDP in this particular case the expression:
\begin{eqnarray}
V(M)\equiv \Omega^{ren}(M;\nu)\Big |_{\nu=0}= \frac{M^2}{\pi
g}+\frac{2M^{3}} {3\pi}.\label{25}
\end{eqnarray}
It follows from (\ref{25}) that at $g>0$, i.e. at $G<G_c$
(\ref{18}), the global minimum point (GMP) of the TDP is arranged at
$M=0$, so the chiral symmetry is not broken. However, at $g<0$, i.e.
at $G>G_c$, the GMP lies at the point $M_0=-1/g$, and there is a
spontaneous chiral symmetry breaking.

Now, let us obtain an alternative expression for the renormalized
TDP (\ref{19}) at $\nu\ne 0$. For this purpose one can rewrite the unrenormalized TDP
$\Omega^{un}(M;\nu)$ (\ref{13}) in the following way
\begin{eqnarray}
\Omega^{un} (M;\nu)&=& \frac{M^2}{4G}-2\int\frac{d^2p}{(2\pi)^2}
\left (2E\right ) -2\int\frac{d^2p}{(2\pi)^2}\Big
(|E+\nu|+|E-\nu|-2E\Big ). \label{013}
\end{eqnarray}
Since the leading terms of the asymptotic expansion (\ref{16}) do
not depend on $\nu$, it is clear that the last integral in
(\ref{013}) is a convergent one. Other terms in (\ref{013}) form the
unrenormalized TDP of the particular case with $\nu=0$ which is
reduced after renormalization procedure to the expression
(\ref{25}). Hence, after renormalization we obtain from (\ref{013})
the following finite expression (evidently, it coincides with
renormalized TDP (\ref{19})):
 \begin{eqnarray}
\Omega^{ren} (M;\nu)=V(M) -2\int\frac{d^2p}{(2\pi)^2}\Big
(|E+\nu|+|E-\nu|-2E \Big ), \label{24}
\end{eqnarray}
where $V(M)$ is presented in (\ref{25}). The integral terms in
(\ref{24}) can be explicitly calculated. As a result, we have
\begin{eqnarray}
\Omega^{ren} (M;\nu)&=&V(M) -\frac{1}{3\pi}\theta
(\nu-M)(\nu-M)^2(2M+\nu). \label{23}
\end{eqnarray}

\subsection{The particular case  $\nu=0$, $B_\perp\ne 0$}
\label{Bperp}

Here we briefly discuss how the nonzero perpendicular magnetic field
$B_\perp$ influences the phase structure of the initial model (1)
in the simplified case of $\mu_B=0$, i.e. when Zeeman interaction is
not taken into account ($\nu=0$). The renormalized TDP of the GN
model with single $O(N)$ fundamental multiplet of four-component
Dirac spinor fields was obtained and investigated in this case in,
e.g., \cite{klimenko3,klimenko,Semenoff:1998bk,zkke}. The
generalization to the case of  GN model with two $O(N)$ multiplets,
as in the model under consideration, is trivial. So we have
\begin{eqnarray}
\Omega^{ren} (M;B_\perp)&=&\frac{M^2}{\pi g}+\frac{MeB_\perp}{\pi}
-\frac{(2eB_\perp)^{3/2}}{\pi }\zeta\left (-\frac
12,\frac{M^2}{2eB_\perp}\right ),
 \label{230}
\end{eqnarray}
where $\zeta(s,x)$ is the generalized Riemann zeta-function. (Since
 $\zeta(-1/2,x)=-2x^{3/2}/3+{\cal
O}(\sqrt{x})$ at $x\to\infty$, it is clear that
 the TDP (\ref{230}) coincides with $V(M)$ (\ref{25}) at $B_\perp=0$.)
It is evident that
\begin{eqnarray}
\frac{\partial \Omega^{ren} (M;B_\perp)}{\partial M}&=&\frac{2M}{\pi
g}+ \frac{eB_\perp}{\pi} -\frac{M(2eB_\perp)^{1/2}}{\pi }\zeta\left
(\frac 12,\frac{M^2}{2eB_\perp}\right ),
 \label{231}
\end{eqnarray}
where we have used the relation $\partial\zeta (s,x)/\partial
x=-s\zeta(s+1,x)$. Since $\zeta(1/2,x)=x^{-1/2}+{\cal O}(x^0)$ at
$x\to 0$, we see from (\ref{231}) that \footnote{Without loss of
generality one can suppose that $eB_\perp\ge 0$.}
\begin{eqnarray}
\frac{\partial \Omega^{ren} (M;B_\perp)}{\partial M}\Big|_{M\to 0_+}&=&-\frac{eB_\perp}{\pi}<0,
 \label{232}
\end{eqnarray}
which means that the TDP (\ref{230}) can never have a global minimum
point at $M=0$, if $B_\perp\ne 0$. Hence, if the model (1) is
subjected to the external (arbitrarily small) perpendicular magnetic
field $B_\perp$ and, in addition, the Zeeman interaction of
electrons with this magnetic field is not taken into account ($\nu=0$), then at arbitrary (even infinitesimal) positive
values of the bare coupling constant $G$ the chiral
symmetry breaking occurs in the model. It
is the so-called magnetic catalysis effect \cite{klimenko}. In
particular, it means that at $g>0$, i.e. at $G<G_c$, the spontaneous
chiral symmetry breaking is induced by external magnetic field. At
$g<0$, i.e. at $G>G_c$, the chiral symmetry is broken even at $\vec
B =0$ due to a rather strong self-interaction coupling constant $G$,
however an external magnetic field enhances chiral symmetry breaking
in this case. Moreover, using in (\ref{231}) the above asymptotic
expansion of the $\zeta(1/2,x)$ function at $x\to 0$, it is possible
to show that at $g>0$ the solution $M_0(B_\perp)$ of the gap
equation $\partial \Omega^{ren} (M;B_\perp)/\partial M =0$ has the
following behavior at $eg^2B_\perp\to 0$ \cite{klimenk,klimenko3}:

\begin{eqnarray}
M_0(B_\perp)=eB_\perp g/2 + {\it o} (egB_\perp).
 \label{233}
\end{eqnarray}
Note, the gap $M_0(B_\perp)\sim\sqrt{eB_\perp}$ at
$eg^2B_\perp\to\infty$ both for negative and positive values of the
coupling $g$ \cite{klimenko3}.

\subsection{The TDP in the general case  $\nu\ne 0$, $B_\perp\ne 0$}

The TDP of the GN model with single $O(N)$ multiplet of Dirac
spinors and at nonzero values of a chemical potential and $B_\perp$ was
obtained, e.g., in \cite{klimenko3,klimenk}. Taking into account the
fact that in our case each of two $O(N)$ multiplets has its own
chemical potential $\mu_k=\pm\nu$, one can easily generalize the
results of \cite{klimenko3,klimenk} and find the following
expression for the renormalized TDP of the GN model (1):
\begin{eqnarray}
\Omega^{ren} (M;\nu,B_\perp)&=&\Omega^{ren} (M;B_\perp)
-\frac{eB_\perp}{\pi}\sum_{n=0}^\infty s_n\theta
(\nu-\varepsilon_n)(\nu-\varepsilon_n),
 \label{2300}
\end{eqnarray}
where $s_n=2-\delta_{0n}$, $\varepsilon_n=\sqrt{M^2+2neB_\perp}$,
and the TDP $\Omega^{ren} (M;B_\perp)$ is presented in (\ref{230}).
Note that due to a presence of the $\theta(x)$-functions, the
summation over $n$ in (\ref{2300}) is performed really from 0 to
Int$[(\nu^2-M^2)/2eB_\perp]$, where Int$(x)$ is the integer part of
$x$. It follows from (\ref{2300}) and (\ref{231}) that the global
minimum point $M_0(B_\perp,\nu)$ (or the gap) of the TDP
(\ref{2300}) obeys the gap equation
\begin{eqnarray}
\frac{\partial \Omega^{ren} (M;\nu,B_\perp)}{\partial
M}&=&\frac{2M}{\pi g}+\frac{eB_\perp}{\pi}
-\frac{M(2eB_\perp)^{1/2}}{\pi }\zeta\left (\frac
12,\frac{M^2}{2eB_\perp}\right
)+\frac{eB_\perp}{\pi}\sum_{n=0}^\infty s_n\theta
(\nu-\varepsilon_n)\frac{M}{\varepsilon_n}=0.
 \label{2301}
\end{eqnarray}
Recall, in the framework of the model (1) the gap $M_0(B_\perp,\nu)$
is also a dynamical mass of quasi-particles (electrons).

\section{Some properties of the model at $g>0$}

Let us first investigate how the external magnetic field influences
the phase structure and magnetization of the model at $g>0$, i.e.
in the case of subcritical values of the bare coupling constant,
$G<G_c$. So, it is necessary to study the properties of the global
minimum point of the TDP (\ref{2300}) at $g>0$. Recall, there
$\nu\equiv\mu_B|\vec B|$, where $|\vec
B|=\sqrt{B_\perp^2+B^2_\parallel}$.

\subsection{Magnetic catalysis effect}

Suppose for a moment that $\nu$ does not depend on $|\vec B|$. In
such a situation the parameter $\nu$ is not the Zeeman splitting
energy, but rather a usual chemical potential. In this case, i.e. when there are two independent external parameters $B_\perp$ and $\nu$, the $(egB_\perp,\nu)$-phase structure of our model (1) is the same as that of the GN model considered in \cite{klimenk}. \footnote{In \cite{klimenk} the GN model with one $O(N)$ multiplet of fermion fields was investigated at $G<G_c$ and in the presence of $B_\perp\ne 0$ and nonzero chemical potential $\nu$. Since the TDP (\ref{2300}) of our model (1) coinñides, up to a factor 2, with the TDP of \cite{klimenk}, it can be concluded that both models have an identical $(egB_\perp,\nu)$-phase structure, if $B_\perp$ and $\nu$ are independent. } Then, as it follows from \cite{klimenk}, for each fixed $B_\perp$
there exists a critical value $\nu_c(B_\perp)$ of the chemical potential
$\nu$ such that at $\nu<\nu_c(B_\perp)$ (at $\nu>\nu_c(B_\perp)$) a
chiral symmetry broken phase (chirally symmetric phase) is realized
in the $(egB_\perp,\nu)$-phase portrait. On the critical curve
$\nu=\nu_c(B_\perp)$ the first order phase transitions take place in
the model. Moreover, it was established in \cite{klimenk} that
$\nu_c(B_\perp)=M_0(B_\perp)+o(egB_\perp)$ at $B_\perp\to 0$ and
$\nu_c(B_\perp)\sim\sqrt{eB_\perp}$ at $B_\perp\to\infty$ ($M_0(B_\perp)$ is the gap (\ref{233})). Hence, and it is the
most important thing for our further consideration, there is a
straight line $\lambda$ in the $(egB_\perp,\nu)$-plane, tangent to a
critical curve $\nu=\nu_c(B_\perp)$ at the point $B_\perp=0$, such
that the whole $(egB_\perp,\nu)$-region above $\lambda$ belongs to a
symmetric phase of the model. It is clear from (\ref{233}) that
\begin{eqnarray}
\lambda=\{(egB_\perp,\nu):~\nu=egB_\perp/2\}.
 \label{26}
\end{eqnarray}
Moreover, any straight line $\nu=kegB_\perp$ with $k<1/2$ crosses
the region of the $(egB_\perp,\nu)$-plane, corresponding to a chiral symmetry broken phase.

\underline{\bf The case $B_\parallel=0$, i.e. $B_\perp=|\vec B|$.}
Now, as it was intended from the very beginning, we suppose that
$\vec B$  and $\nu$ are dependent quantities and, furthermore, that
the external magnetic field $\vec B$ is perpendicular to a system
plane, i.e. $B_\perp=|\vec B|$ and $\nu =\mu_BB_\perp$. Hence, in
the case under consideration only the points of the straight line
$\nu =\mu_BB_\perp\equiv\kappa egB_\perp$ of the above mentioned
$(egB_\perp,\nu)$-plane are relevant to a real physical situation
(evidently, $\kappa =\mu_B/(eg)$). So, if $\kappa>1/2$, i.e. at
sufficiently small values of $g$, then the straight line $\nu
=\mu_BB_\perp$ as a whole is above the line $\lambda$ (\ref{26}),
and spontaneous chiral symmetry breaking is forbidden in the system.
However, if the coupling constant $g$ is greater than
$g_c=2\mu_B/e$, we have $\kappa<1/2$ and the line $\nu
=\mu_BB_\perp$ is below $\lambda$. Obviously, in this case the
straight line $\nu =\mu_BB_\perp$ crosses the region of the
$(egB_\perp,\nu)$-plane with chiral symmetry breaking. Hence, at
$g>g_c$ chiral symmetry might be broken only for some finite
interval of $B_\perp$-values. It means that the magnetic catalysis
effect at $B_\parallel=0$ and $\mu_B\ne 0$, i.e. when the Zeeman
interaction of electrons with magnetic field is taken into account,
is qualitatively different from the case with $B_\parallel=0$ and
$\mu_B=0$ (see the section \ref{Bperp}). Indeed, i) at $\mu_B=0$ the
external (arbitrary small) magnetic field $B_\perp$ induces
spontaneous chiral symmetry breaking at arbitrary values of $g>0$
(see the section \ref{Bperp}), whereas at $\mu_B\ne 0$ chiral
symmetry might be broken by $B_\perp$ only at $g>g_c>0$. ii) If
$g>g_c$, then at $\mu_B\ne 0$ the chiral symmetry is allowed to be
spontaneously broken only for rather small values of $B_\perp$,
i.e. at $B_\perp<B_{\perp c}$, where $0<B_{\perp c}<\infty$.  The
symmetry is restored at sufficiently high  values of
$B_\perp>B_{\perp c}$. In contrast, if the Zeeman interaction is
neglected, we have $B_{\perp c}=\infty$ for arbitrary $g>0$.

To illustrate these circumstances we made some numerical
investigations of the TDP (\ref{2300}) at $B_\perp=|\vec B|$. For
example, we have found that at $g=2.5g_c$, $g=3.5g_c$ and $g=5g_c$
the corresponding critical values $B_{\perp c}$ of the perpendicular
magnetic field at which there is a restoration of the chiral
symmetry  are the following, $eg^2B_{\perp c}\approx 0.059$,
$eg^2B_{\perp c}\approx 0.518$ and $eg^2B_{\perp c}\approx 2.04$.
Moreover, the behavior of the dynamical electron mass (or the gap)
$M_0(B_\perp,\nu)$ vs $B_\perp$ in the particular case $g=5g_c$ is
presented in Fig. 1. It is clear from this figure that the gap is an
increasing function vs $B_\perp$ up to a critical value $B_{\perp
c}$, where it vanishes sharply, i.e. the first order phase
transition occurs.

\underline{\bf The case $B_\perp\ne |\vec B|$.} Now let us consider
the general case when $B_\parallel\ne 0$, i.e. $B_\perp\ne |\vec
B|$. In this case the mass gap $M_0(B_\perp,\nu)$ is really a
function of two independent quantities, $B_\perp$ and $|\vec B|$,
with an additional evident physical constraint $B_\perp\le |\vec B|$.
Investigating properties of the global minimum point of the TDP
(\ref{2300}), depending on $B_\perp$ and $|\vec B|$, it is possible
to obtain a corresponding phase portrait of the model. For a typical
value of the parameter $g=5g_c$ the phase structure of the model is
presented in Fig. 2.

It is clear from the figure that at arbitrary small and
perpendicular external magnetic field $\vec B$, such that $|\vec
B|<B_{\perp c}$ (see the previous paragraphs), the system is in the
chiral symmetry broken phase 2. Then, the chiral symmetry can be restored
by two qualitatively different ways. First, one may increase the
strength of $\vec B$, or, second, it is possible simply to tilt
$\vec B$ with respect to a system plane. In the last case, not too
high deflection angle $\phi$ of the magnetic field is needed
($\phi\approx 45^o$, where $\phi$ is the angle between $\vec B$ and
the normal to the system plane) in order to restore the
symmetry.
\begin{figure}
%----figure 1,2
\includegraphics[width=0.45\textwidth]{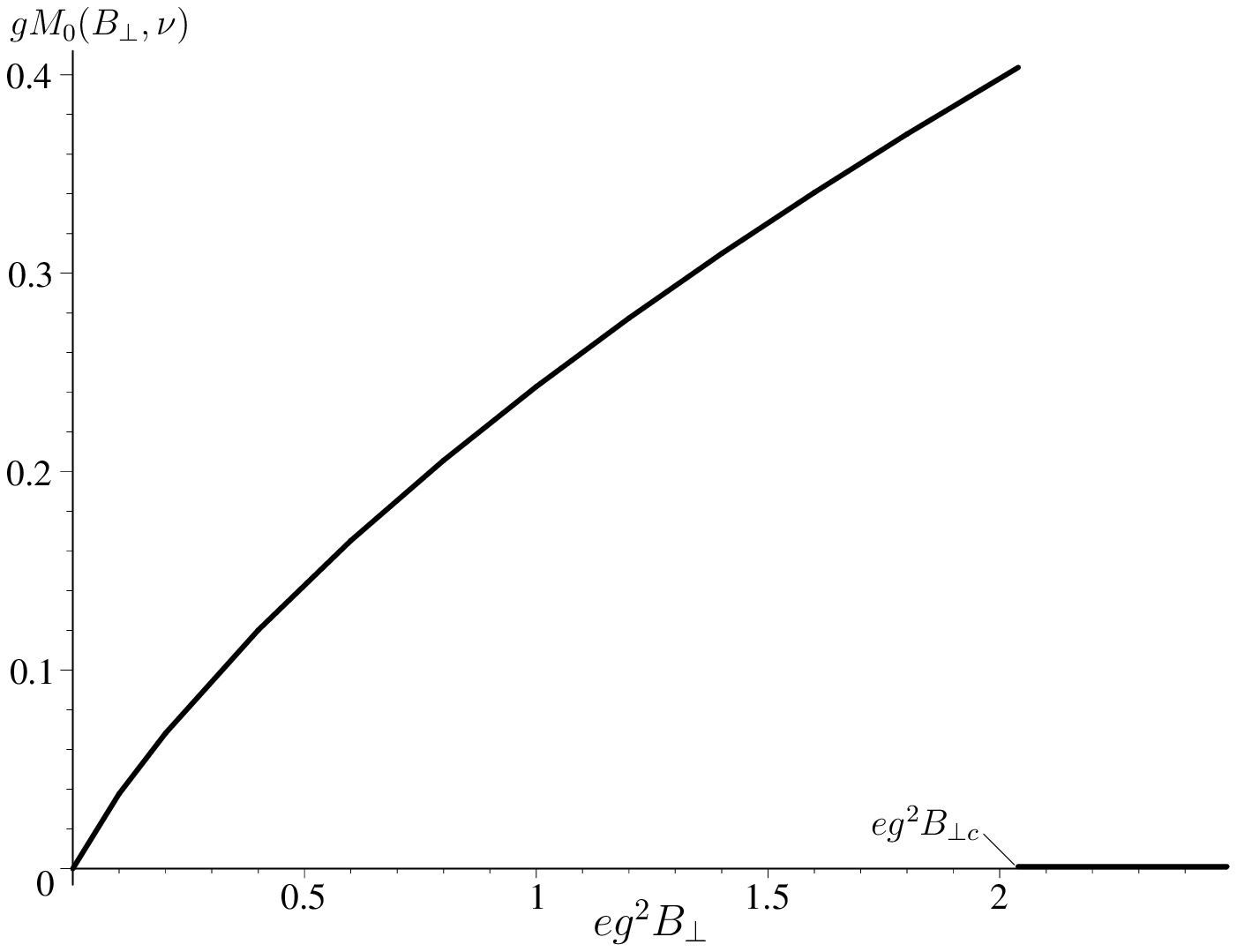}
 \hfill
\includegraphics[width=0.45\textwidth]{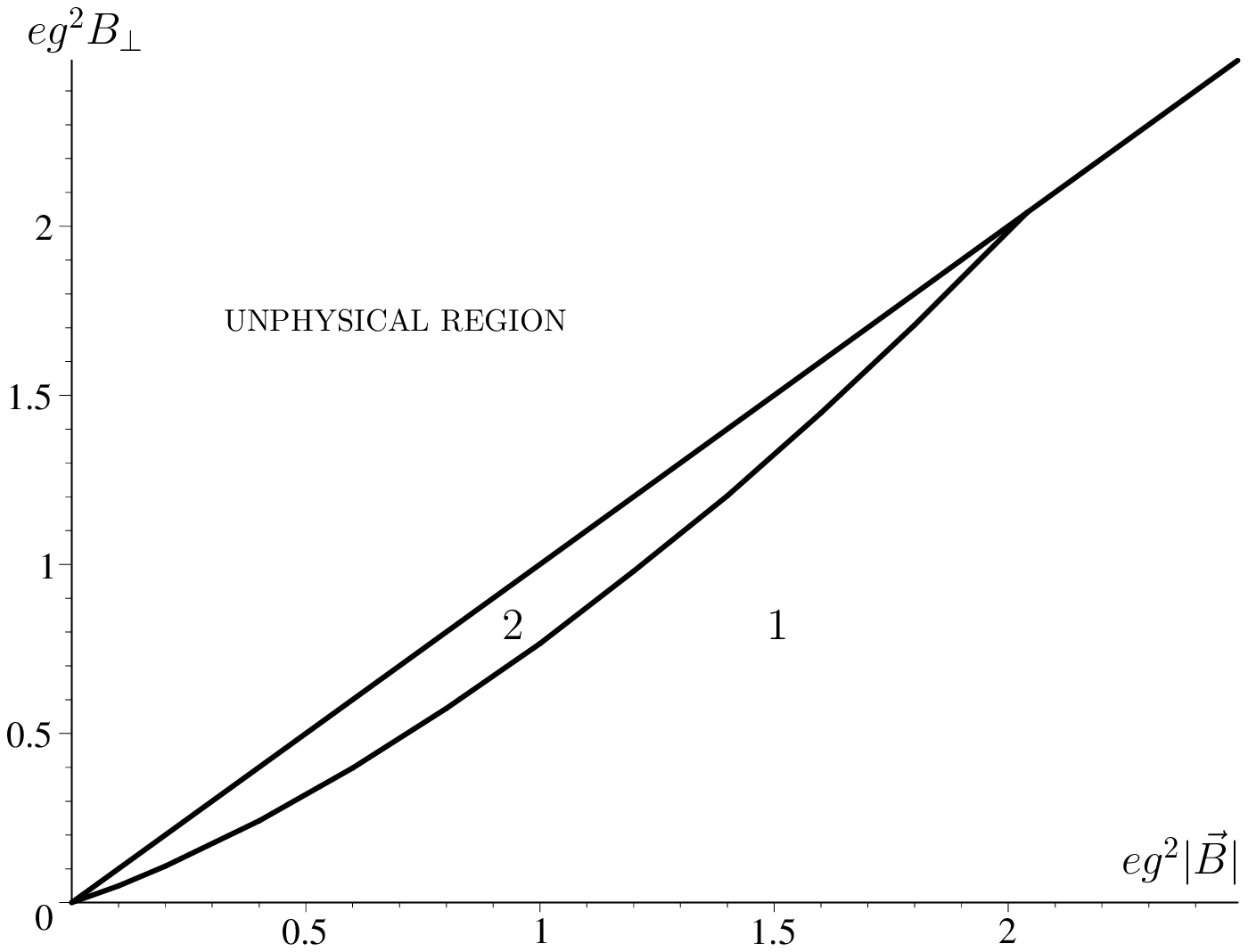}\\
\parbox[t]{0.45\textwidth}{
 \caption{{\it The case $g>0$}: The mass gap $M_0(B_\perp,\nu)$ vs $B_\perp$ in the particular case
$B_\parallel=0$ and $g=5g_c\equiv 10\mu_B/e$. Here $eg^2B_{\perp
c}\approx 2.04$. }
 }\hfill
\parbox[t]{0.45\textwidth}{
\caption{{\it The case $g>0$}: The $(|\vec B|,B_\perp)$-phase
portrait of the model at $g=5g_c\equiv 10\mu_B/e$. The numbers 1 and 2 denote the chirally symmetric and chirally broken phases, respectively. In the unphysical region of the figure $B_\perp>|\vec B|$. The boundary
between 1 and 2 phases is the curve of the first order phase
transitions. } }
\end{figure}
\begin{figure}
%----figure 3,4
\includegraphics[width=0.45\textwidth]{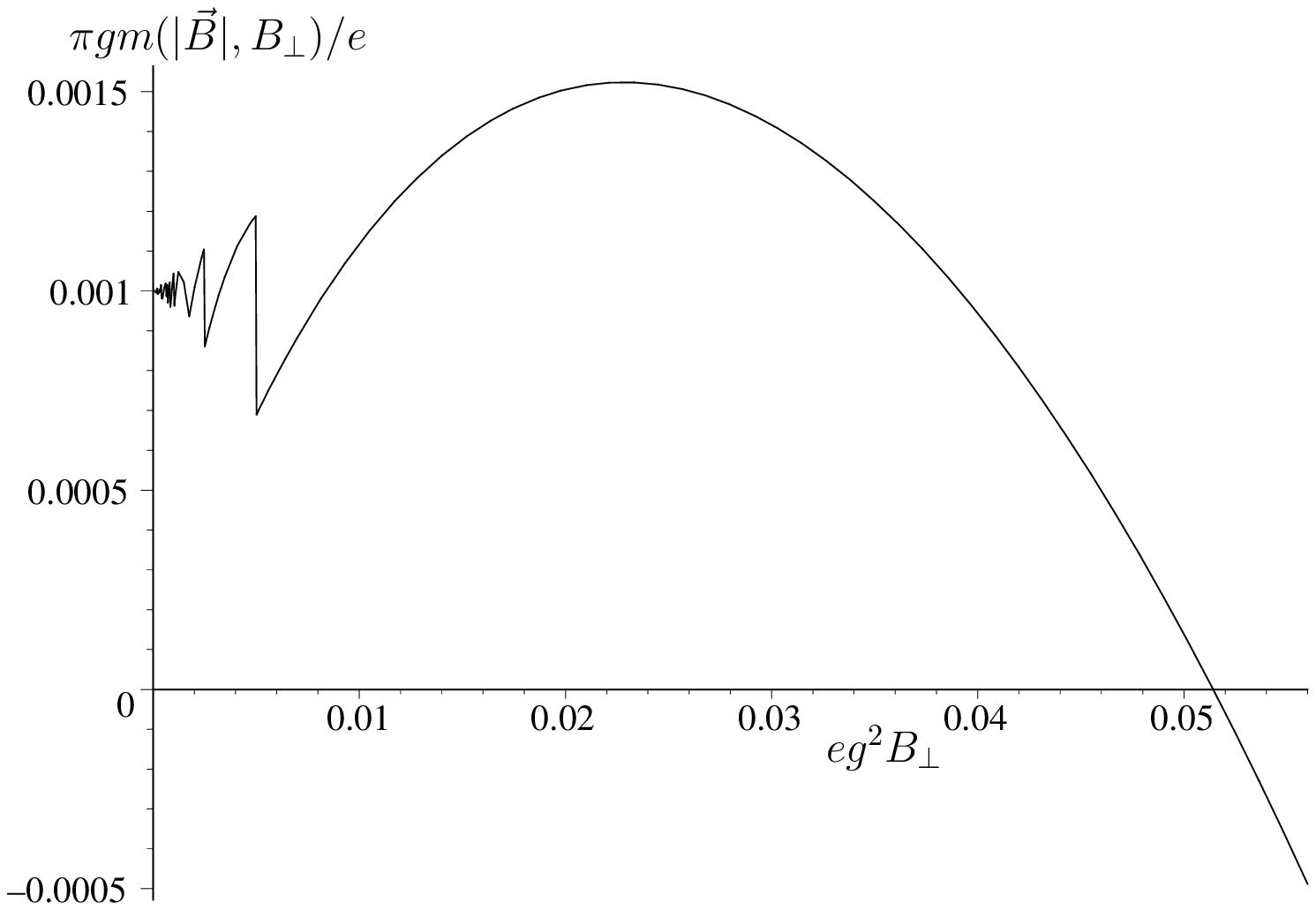}
 \hfill
\includegraphics[width=0.45\textwidth]{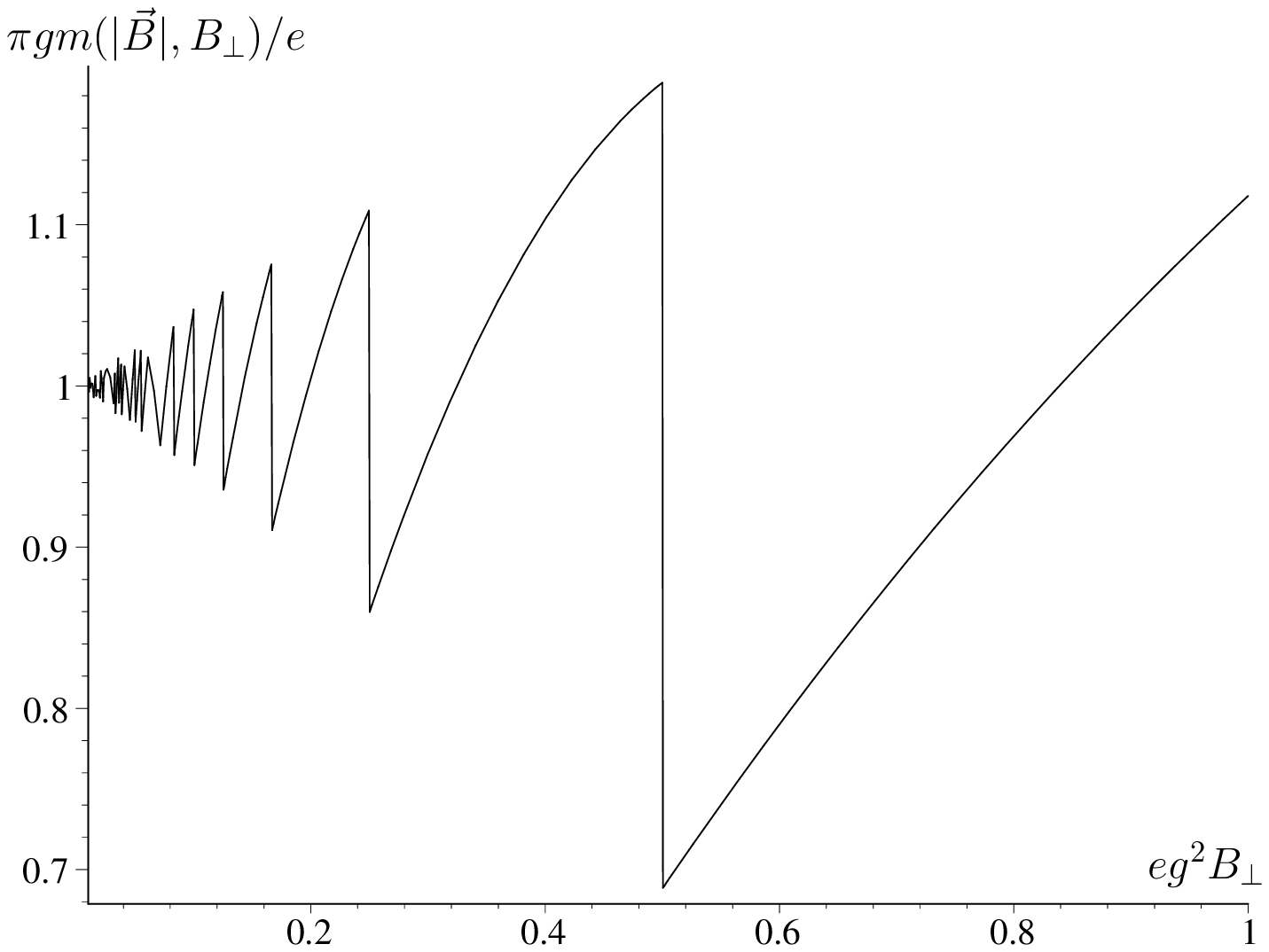}\\
\parbox[t]{0.45\textwidth}{
 \caption{{\it The case $g>0$}: Magnetization $m(|\vec
B|,B_\perp)$ vs $B_\perp$ at fixed $eg^2|\vec B|=1$ and
$g=5g_c\equiv 10\mu_B/e$.  }
 }\hfill
\parbox[t]{0.45\textwidth}{
\caption{{\it The case $g>0$}: Magnetization $m(|\vec B|,B_\perp)$
vs $B_\perp$ at fixed $eg^2|\vec B|=1$ and $g=0.5g_c\equiv
\mu_B/e$.} }
\end{figure}

\subsection{Oscillations of the magnetization}

Now, let us consider the magnetization $m(|\vec B|,B_\perp)$ of the
system under influence of an external tilted magnetic field at
$g>0$. At fixed angle $\phi$ between $\vec B$ and the normal to the
system plane, we define the magnetization by the following relation
\begin{eqnarray}
m(|\vec B|,B_\perp)\equiv -\frac{d\Omega^{ren}
(M;\nu,B_\perp)}{d|\vec B|}\Big|_{M=M_0(B_\perp,\nu)},
 \label{27}
\end{eqnarray}
where $M_0(B_\perp,\nu)$ is the mass gap. Certainly, one should take
into account that at fixed $\phi$ the perpendicular component
$B_\perp$ of the magnetic field is proportional to $|\vec B|$, i.e.
$B_\perp=|\vec B|\cos\phi$, so in (\ref{27}) we have
\begin{eqnarray}
\frac{d}{d|\vec B|}=\frac{\partial}{\partial |\vec
B|}+\frac{B_\perp}{|\vec B|}\frac{\partial}{\partial B_\perp}.
 \label{270}
\end{eqnarray}
 Taking this relation into account, it is possible to obtain
\begin{eqnarray}
m(|\vec B|,B_\perp)=-\frac{B_\perp}{|\vec
B|}\frac{\partial\Omega^{ren} (M;B_\perp)}{\partial
B_\perp}\Bigg|_{M=M_0(B_\perp,\nu)}+\frac{eB_\perp}{\pi |\vec
B|}\sum_{n=0}^\infty s_n\theta (\nu-\varepsilon_n)\left
(2\nu-\frac{\varepsilon_n^2+enB_\perp}{\varepsilon_n}\right
)\Bigg|_{M=M_0(B_\perp,\nu)},
 \label{28}
\end{eqnarray}
where the notations of the expression (\ref{2300}) are used. In the
chirally broken phase 2 of Fig. 2 the mass gap $M_0(B_\perp,\nu)$ is
greater than $\nu$, so the series in (\ref{28}) does not contribute to
the magnetization. As a result, the magnetization $m(|\vec
B|,B_\perp)$ is a rather smooth function both over $|\vec B|$ and
$B_\perp$ in the phase 2. However, in the chirally symmetric phase 1
of Fig. 2 we have $M_0(B_\perp,\nu)\equiv 0$ and, as a result, a
jumping and discontinuous (or oscillating) behavior of the
magnetization, which is originated just due to contributions coming
from the series terms in (\ref{28}). Indeed, at $M_0(B_\perp,\nu)=
0$ we have for the magnetization in the chirally symmetric phase 1:
\begin{eqnarray}
m(|\vec B|,B_\perp)\Big |_{\mbox{phase 1}}&=&\frac{eB_\perp}{\pi
}\left [\frac{3}{|\vec B|}\sqrt{2eB_\perp}\zeta (-1/2)+2\mu_B\right ]+\nonumber\\
&&\frac{2eB_\perp}{\pi |\vec B|}\sum_{n=1}^\infty \theta
(\nu-\sqrt{2enB_\perp})\left
(2\nu-\frac{3}{2}\sqrt{2enB_\perp}\right ),
 \label{29}
\end{eqnarray}
where $\zeta (-1/2)\approx -0.208$. The plot of the function
(\ref{29}) $m(|\vec B|,B_\perp)$ vs $B_\perp$ is presented in Figs 3
and 4 in two particular cases $g=5g_c$ and $g=0.5g_c$,
correspondingly, at fixed value of $|\vec B|$ such that $eg^2|\vec
B|=1$. It is clear from these figures that in the region of small
values of $B_\perp$ the quantity (\ref{29}) is a highly oscillating
function.

Suppose that $|\vec B|$ is fixed. Since all terms of the series in
(\ref{29}) are positive quantities, one can conclude that in the
region of sufficiently small $B_\perp$ both the expression in the
square brackets of (\ref{29}) and the magnetization as a whole are
also positive quantities. Hence, at small values of $B_\perp$ the
ground state of the model is a paramagnetic one. The situation can
be changed, if $B_\perp$ approaches $|\vec B|$. In this case,
depending on the relation between dimensionless parameters $e$ and
$\mu_B/g$, one can obtain quite different magnetic properties of the
ground state. Really, if $\mu_B/g\ge e$ (see, e.g., Fig. 4), then
the magnetization is positive for all physical values of $B_\perp$,
$0\le B_\perp\le |\vec B|$, and the system is in the paramagnetic
ground state. However, for a sufficiently small values of
$\mu_B/g\ll e$ there is an interval of rather large values of
$B_\perp$, where both the expression in the square brackets of
(\ref{29}) and the magnetization $m(|\vec B|,B_\perp)$ are negative
quantities, so we have in this case a diamagnetic ground state of
the system. For example, in Fig. 3 a graph of the magnetization
$m(|\vec B|,B_\perp)$ vs $B_\perp$ is drown at fixed $|\vec B|$ and
at $\mu_B/g= 0.1 e$. Clearly, in this case the system is in the
paramagnetic state if $eg^2B_\perp< 0.051$, and it is a diamagnetic
one at $eg^2B_\perp > 0.051$.

Clearly, in the range of $B_\perp$ values, which is near $|\vec B|$,
the series in (\ref{29}) consists of a finite number of nonzero terms
(the greater the value of $B_\perp$, the less the number of nonzero terms there). So, the behavior of the magnetization vs $B_\perp$ can
be easily determined in this case. However, it is difficult to get
any information about the low $B_\perp$ behavior of the
magnetization just from the expression (\ref{29}), since at
$B_\perp\to 0$ an infinite number of nonzero terms appears in the
sum of (\ref{29}). The problem is well known in solid state physics
\cite{osc,osc2} as well as in relativistic condensed matter systems
\cite{elmfors,vshivtsev}, where the magnetic oscillations were
investigated. Of course, the corresponding techniques can be used in
the present model as well.

To study the $B_\perp\to 0$ asymptotic behavior of the magnetization we
apply in (\ref{29}) the well-known Poisson summation formula
\cite{osc2}
\begin{eqnarray}
\sum^{\infty}_{n=0}\alpha_n\Phi (n) =2\sum^{\infty}_{k=0}\alpha_k
\int\limits^{\infty}_{0}\Phi (x) \cos (2\pi kx)dx,
\end{eqnarray}
where $\alpha_k=2-\delta_{0k}$. Then, after rather tedious
calculations  it is possible to find the following asymptotic
behavior of the magnetization (\ref{29}) at $\vec B_\perp\to 0$ and
arbitrary fixed $|\vec B|$ (recall, $\nu=\mu_B|\vec B|$):
\begin{eqnarray}
m(|\vec B|,B_\perp)=\frac{\mu_B\nu^2}{\pi}+\frac{\mu_B
eB_\perp}{\pi^2}\sum^{\infty}_{n=1}\frac 1k \sin\left (\frac{\pi
k}{eB_\perp}\nu^2\right )+o(eB_\perp). \label{30}
\end{eqnarray}
Remark, the leading asymptotic term in this expression, i.e. the
first term in the right hand side of (\ref{30}), is the
magnetization corresponding to the TDP (\ref{23}) with zero
$B_\perp$ component of an external magnetic field. Moreover, an
infinite series in (\ref{30}) is no more than Fourier expansion of
the periodic function $f(x)$, where $x=\nu^2/(2eB_\perp)$. Its
period is equal to unity and in the interval $0<x<1$ it looks like
$f(x)=\pi/2-\pi x$.

Note, in condensed matter systems, both nonrelativistic
\cite{osc,osc2} and relativistic \cite{elmfors,vshivtsev,sadooghi}, magnetic
oscillations usually occur in the presence of chemical potential
$\mu$, i.e. in the systems with $\mu=0$ magnetic oscillations are
absent as a rule. However, as it follows from our consideration (see
also a more earlier paper \cite{schakel}), in systems with planar
structure magnetic oscillations can be induced even at $\mu=0$  by
tilting the external magnetic field with respect to a system plane.

\section{Phase structure of the model at $g<0$ }

In the present section we study the influence of an external
magnetic field on the properties of the initial model (1) at $g<0$,
i.e. at supercritical values of the bare coupling constant, $G>G_c$.
Recall, when the Zeeman interaction is not taken into account the
chiral symmetry breaking, induced originally in this case by a
rather strong coupling, is enhanced additionally by external
magnetic field (see, e.g., in \cite{klimenko,Semenoff:1998bk,zkke}).
It means that dynamical mass of electrons is an increasing function
vs $B_\perp$ throughout the interval $0<B_\perp<\infty$ (in this
case $B_\parallel$ does not influence the properties of the
model). It turns out that Zeeman interaction drastically changes
properties of the model.

\subsection{The particular case, $|g|= \mu_B/e$.}

\underline{\bf The case of perpendicular magnetic field.} First, let
us suppose that external magnetic field $\vec B$ is directed
normally to a system plane, i.e. $B_\perp=|\vec B|$ and
$B_\parallel=0$. For simplicity, we fix the value of $g$ by the
relation $|g|= \mu_B/e$. Investigating in this case the TDP
(\ref{2300}) as well as the gap equation (\ref{2301}), we have found
the behavior of the mass gap $M_0(B_\perp,\nu)$ vs $B_\perp$ (it is
the curve 1 in Fig. 5). It turns out that up to a some critical
value $B_{\perp c_1}$ (such that $eg^2B_{\perp c_1}\approx 0.81$)
the enhancement scenario is realized, i.e. the mass gap is an
increasing function vs $B_\perp$. Moreover, in this chirally broken
phase the gap $M_0(B_\perp,\nu)$ takes rather large values, such
that $M_0(B_\perp,\nu)>\nu$. Consequently, the contribution to the
magnetization $m(|\vec B|,B_\perp)$ coming from the Zeeman
interaction vanishes, i.e. all terms of the series in (\ref{28}) are
zero.  As a result, the magnetization in this phase is completely determined by an interaction of $\vec B$ with orbital angular momentum.
Due to this reason $m(|\vec B|,B_\perp)$ is negative at
$0<B_\perp<B_{\perp c_1}$ (see Fig. 5, where the curve 2 corresponds
to a magnetization), and the ground state of this phase is a
diamagnetic one.

Then, in the critical point $B_\perp=B_{\perp c_1}$ the mass gap
$M_0(B_\perp,\nu)$ jumps to a significantly smaller nonzero value,
and there is a phase transition of the first order to another
chirally broken phase. Further increasing of $B_\perp$ leads to a
restoration of the chiral symmetry at $B_\perp=B_{\perp c_2}$, where
$eg^2B_{\perp c_2}\approx 0.94$. It is a second order phase
transition, since in this point the mass gap $M_0(B_\perp,\nu)$
continuously turns into zero (see Fig. 5). Note also that both in
the second chirally broken phase (at $B_{\perp c_1}<B_\perp<B_{\perp
c_2}$) and in the chirally symmetric one (at $B_{\perp
c_2}<B_\perp<\infty$) the magnetization of the system is positive,
i.e. the ground states of these phases are paramagnetic (see Fig.
5).

\underline{\bf The case of tilted magnetic field.} Now, a few words
about a response of the system with $g<0$ upon an arbitrarily directed
external magnetic field, i.e. when $B_\perp\ne |\vec B|$. Numerical
investigations of the TDP (\ref{2300}), where for simplicity we put
$|g|= \mu_B/e$, bring us to the phase portrait of the model
presented in Fig. 6. There the number 1 corresponds to a chirally
symmetric paramagnetic phase, whereas notations 2 and 3 are used for
two different chirally broken phases. The first of them, i.e. the
phase 2, is a diamagnetic with $m(|\vec B|,B_\perp)<0$, however the
second one, i.e. the phase 3, is a phase with paramagnetic ground
state, since in this region $m(|\vec B|,B_\perp)>0$. Note, at $g<0$
one can also observe the oscillations of the magnetization only in
the chirally symmetric phase 1 when $B_\perp\to 0$.

As it is clear from Figs 5 and 6 the presence of the Zeeman
interaction significantly changes the behavior of the chiral
symmetry under influence of an external both perpendicular and
tilted magnetic field at $g<0$. Indeed, at $\mu_B\ne 0$ the
enhancement of a chiral condensation in this case takes place only
at sufficiently small values of $|\vec B|$, i.e. in the phase 2 of
Fig. 6 (it means that fixing the tilting angle of the magnetic field
we obtain the growth of the mass gap $M_0(B_\perp,\nu)$ at
increasing $|\vec B|$).  Further increasing of $|\vec B|$ leads
ultimately to a chiral symmetry restoration.
\begin{figure}
%----figure 5,6
\includegraphics[width=0.45\textwidth]{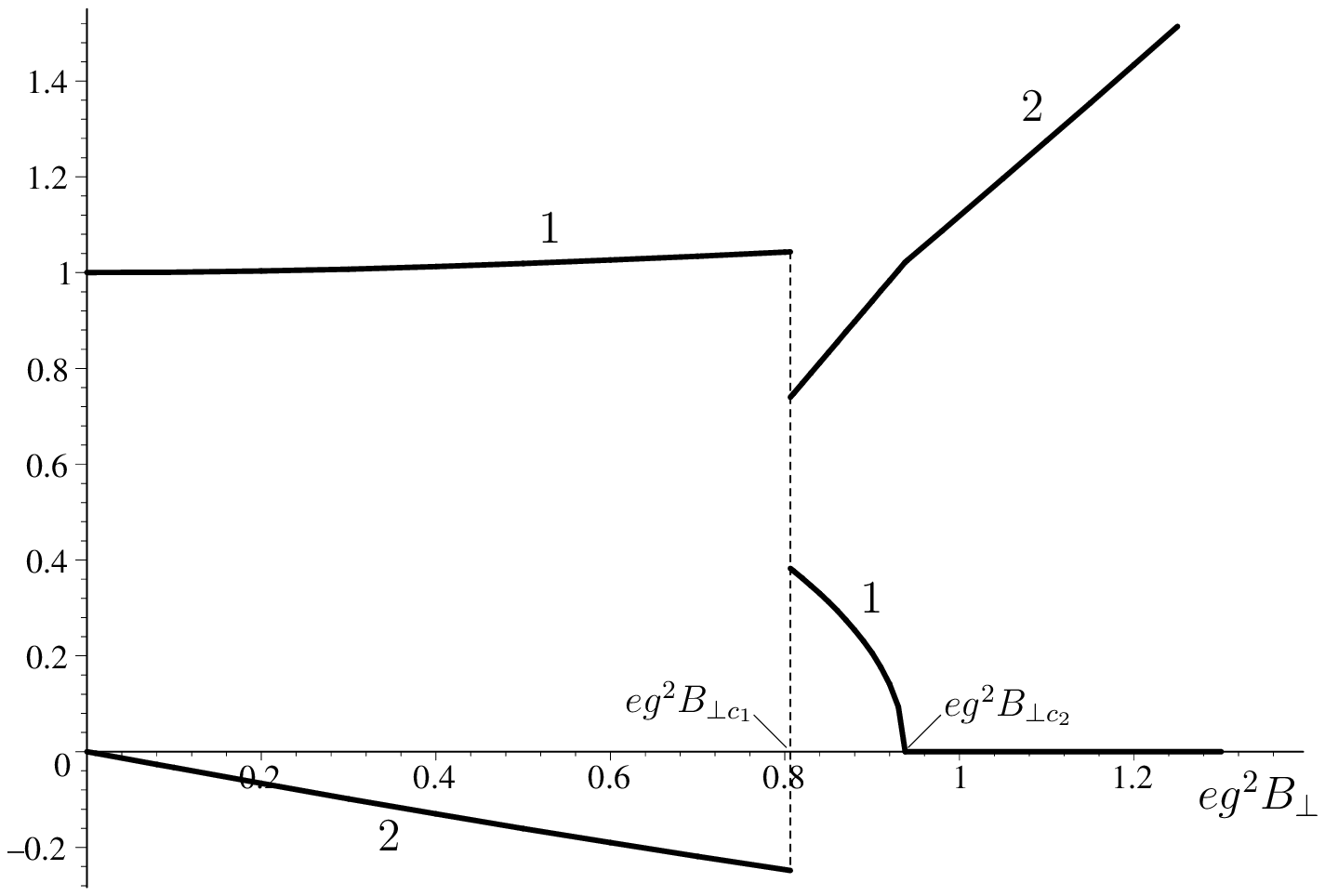}
 \hfill
\includegraphics[width=0.45\textwidth]{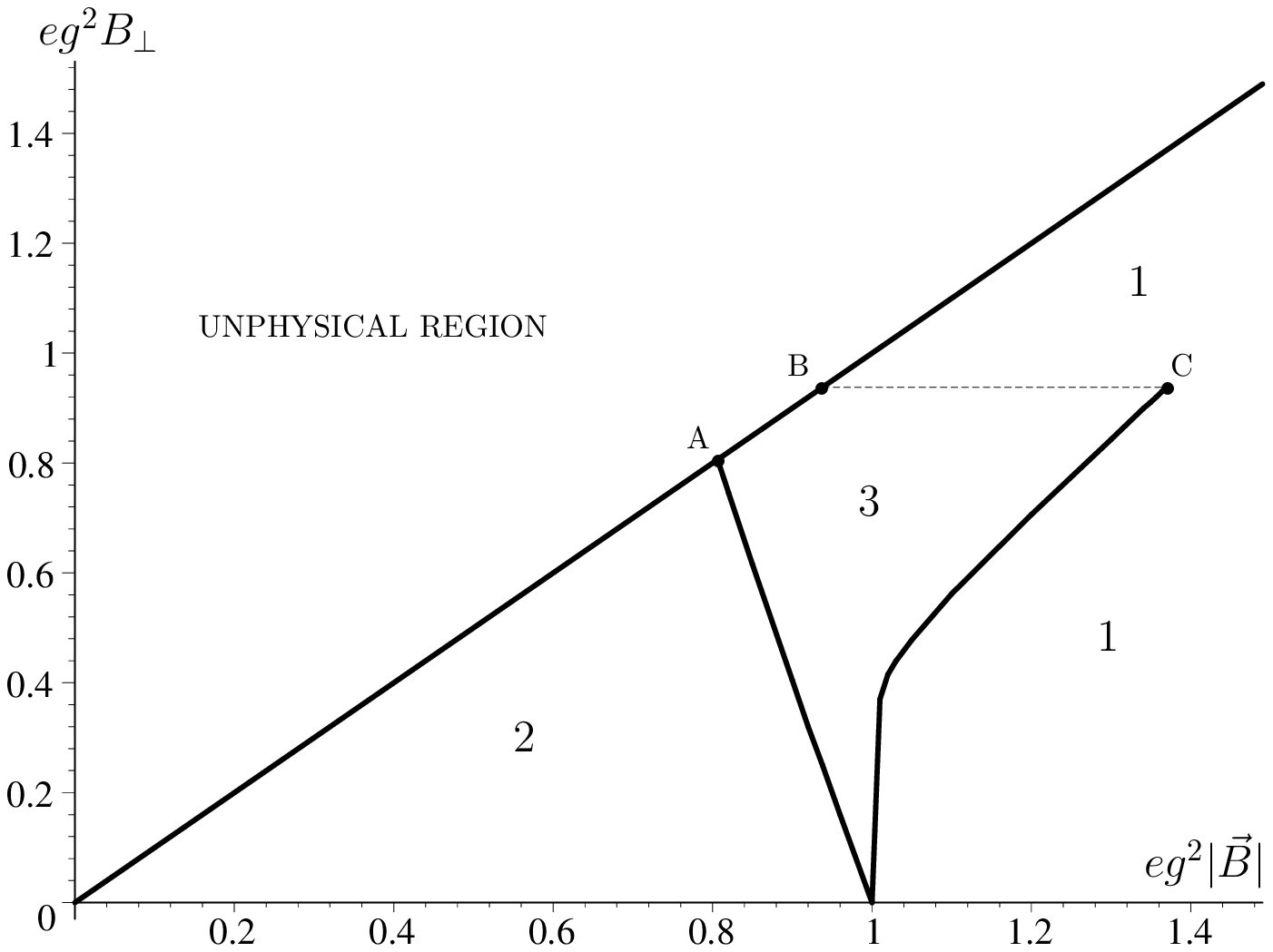}\\
\parbox[t]{0.45\textwidth}{
 \caption{{\it The case $g<0$}: Mass gap $M_0(B_\perp,\nu)$ and magnetization $m(|\vec B|,B_\perp)$
 vs $B_\perp$ in the particular case
$B_\parallel=0$ and $|g|= \mu_B/e$. Curves 1 and 2 are the plots of
the dimensionless quantities $gM_0(B_\perp,\nu)$ and $\pi g m(|\vec
B|,B_\perp)/e$, correspondingly. Here $eg^2B_{\perp c_1}\approx
0.81$ and $eg^2B_{\perp c_2}\approx 0.94$. }
 }\hfill
\parbox[t]{0.45\textwidth}{
\caption{{\it The case $g<0$}: The $(|\vec B|,B_\perp)$-phase
portrait of the model at $|g|= \mu_B/e$. The numbers 1 denote the
chirally symmetric phase, whereas the numbers 2 and 3 denote two
different chirally broken phases (on the boundary between 2 and 3
the mass gap changes by a jump). The coordinates of the points A, B
and C approximately are $(0.81,0.81)$, $(0.94,0.94)$ and
$(1.37,0.94)$, correspondingly. The line BC is a curve of second
order phase transitions; on the other lines the first order phase
transitions take place. The unphysical region of the figure
corresponds to $B_\perp>|\vec B|$.  } }
\end{figure}

\subsection{Phase structure in the general case }

Clearly, for other relations between $|g|$ and $\mu_B$, i.e. at
$|g|\ne \mu_B/e$, the $(eg^2|\vec B|,eg^2B_{\perp})$-phase portrait
of the model might be quite different from Fig. 6. To imagine the
phase structure of the model for an arbitrary, but fixed, relation
between $|g|$ and $\mu_B$ it is very convenient to use for its
description the new dimensionless parameters, $x= \mu_B|\vec B||g|$
and $y=eg^2B_{\perp}$. \footnote{Strictly speaking, only $x$ is a
new parameter, since $y$ was  already used in Fig. 6 etc.} Assuming
for a moment that $x$ and $y$ are fully independent quantities, it
is possible to investigate the behavior of the global minimum point
of the TDP (25) as a function of $x$ and $y$ and then to obtain the
$(x,y)$-phase portrait of the model depicted in Figs 7 and/or 8.
(The line L of these figures should be ignored in this case. Note
also that in Fig. 8 the phase portrait is depicted for a more
extended region of the parameter $y$.) There one can see only three
different phases which were already presented in Fig. 6. So we use
the same notations for them, 1, 2 and 3. In reality, there is a
constraint between $x$ and $y$ which is due to the physical
requirement $B_\perp\le |\vec B|$. In terms of $x$ and $y$ it looks
like $y\le cx$, where $c=e|g|/\mu_B$, i.e. not the whole
$(x,y)$-plates of Figs 7 and 8 can be considered as a phase diagram,
but only those areas which are below the line L. The points of the
line L correspond to a perpendicular external magnetic field, i.e. we have $B_{\perp}=|\vec B|$ on the line L. Clearly, if the quantity
$c=e|g|/\mu_B$ varies, then the line L of Figs 7 and 8 changes its
slope and, as a result, the allowed physical region which is below L
is also changed. However, the positions and forms of the critical
curves in Figs 7, 8 are not changed at different values of the
parameter $c$.

It is easily seen from Fig. 8 that inside the interval  $3<y<11$ the
critical curve $l$ of the phase diagram can be approximated by a
straight line with a slope coefficient $c^*\approx 28$.
Extrapolating this behavior of the curve $l$ to the region with
higher $y$-values, one can conclude that a typical phase portrait of
the initial model corresponding to the weak coupling $|g|$, such that
$c=e|g|/\mu_B<c^*$, is presented in Fig. 7 (it is the region just
below the line L). In this case the line L certainly crosses
critical curve $l$ of a phase portrait, i.e. it passes through
several different phases, including the chirally symmetric phase 1.
As a result, one can see that at $c<c^*$ the chiral symmetry is
always restored at $|\vec B|\to\infty$ irrespective of the magnetic
field directions (even at a perpendicular magnetic field). In
particular, the case $c=1$ was considered in details in the previous
section IV A, and Fig. 7 at $c=1$ coincides with the phase diagram
of Fig. 6.

In contrast, if $c>c^*$ then a typical phase portrait of the model
is depicted in Fig. 8 (it is a region which is below and/or to the
right of the line L). Clearly, in this case the line L does not
cross any of the critical curves of the phase diagram, and at
arbitrary values of a {\it perpendicular} magnetic field the chiral
symmetry cannot be restored, since we move along the line L when
$B_\perp=|\vec B|$ increases. However, if $|\vec B|$ reaches the
values corresponding to $x>0.7$, then in this case at fixed $|\vec
B|$ it is also possible to restore the symmetry by tilting the
magnetic field away from the normal direction. In particular, if the
parameter $x$ lies, e.g., in the interval $0.7<x<1.4$ (see Fig. 8),
then a number of phase transitions can occur in the system that are
also caused only by the inclination of an external magnetic field.
\begin{figure}
%----figure 7,8
\includegraphics[width=0.45\textwidth]{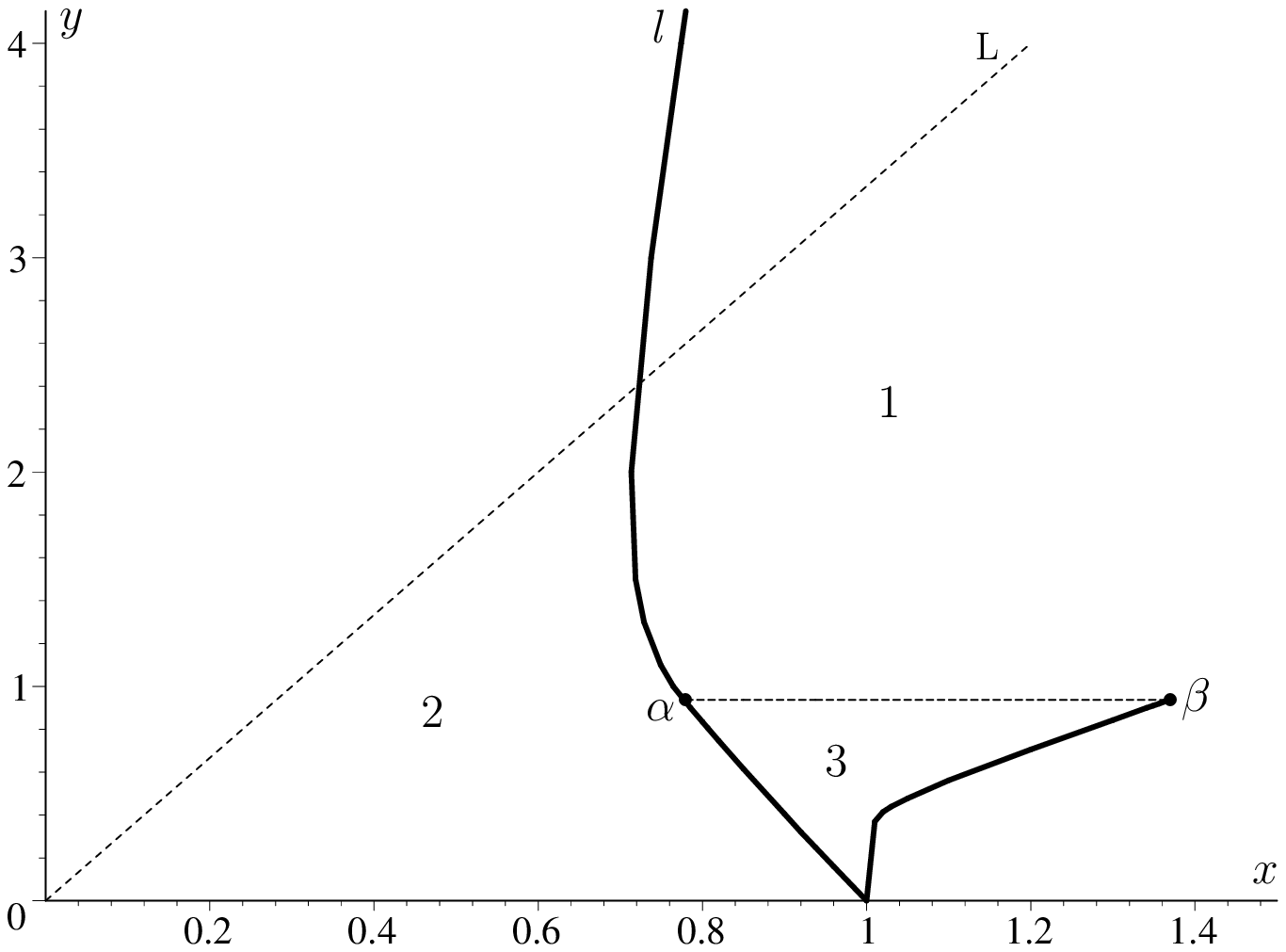}
 \hfill
\includegraphics[width=0.45\textwidth]{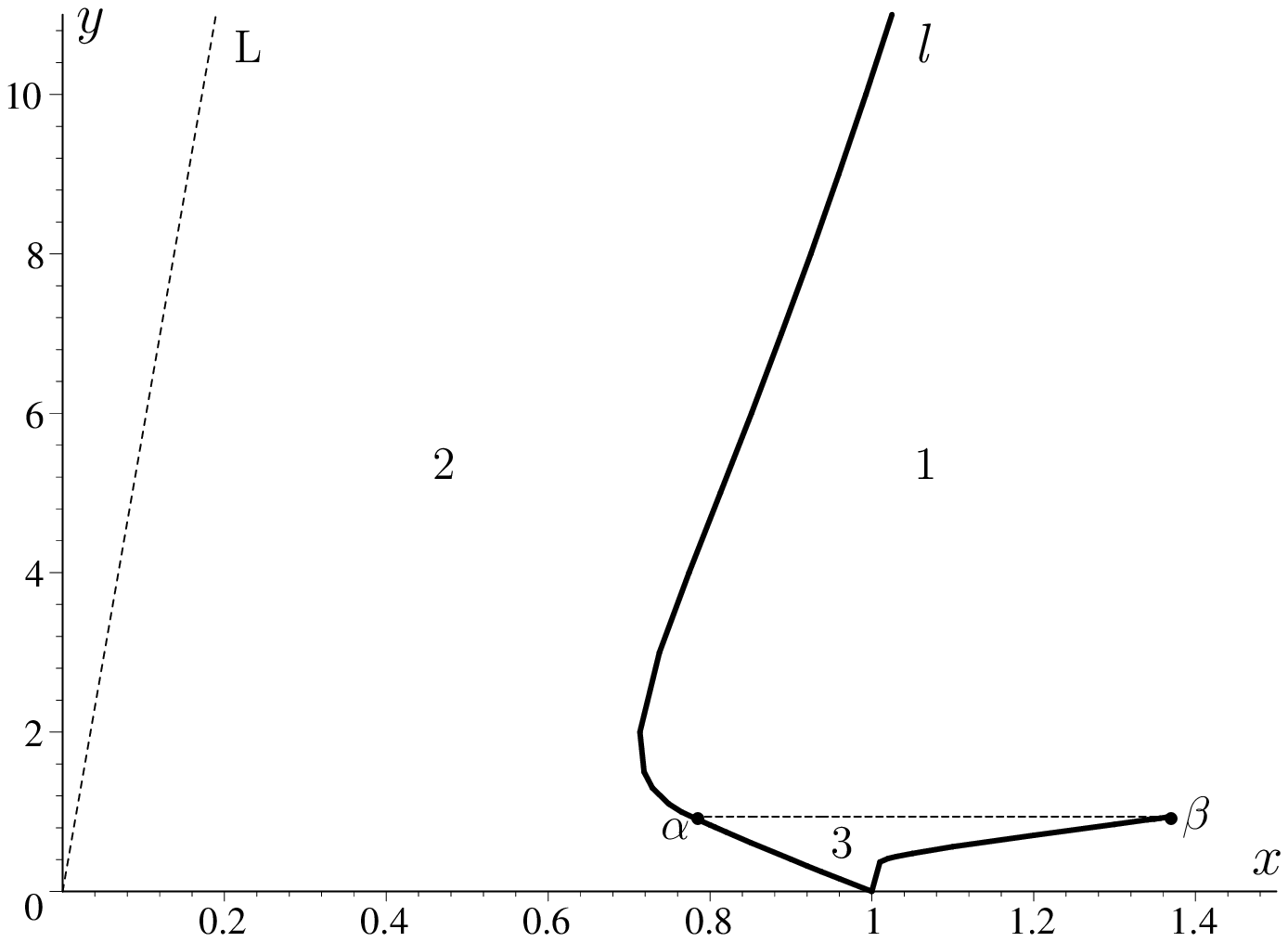}\\
\parbox[t]{0.45\textwidth}{
 \caption{{\it The case $g<0$}: The $(x,y)$-phase diagram of the model, where $x= \mu_B|\vec B||g|$
and $y=eg^2B_{\perp}$, typical for values of $c\equiv
e|g|/\mu_B<c^*\approx 28$. Physical region of the diagram
corresponding to $B_\perp\le |\vec B|$ relation lies just below the
line L=$\{(x,y):~y=cx\}$. The notations 1, 2 and 3 for different
phases of the system are the same as in Fig. 6. First order phase
transitions occur on the solid curves. On the line $\alpha\beta$
second order phase transitions take place. $\alpha\approx
(0.71,0.94)$, $\beta\approx (1.37,0.94)$.}
 }\hfill
\parbox[t]{0.45\textwidth}{
\caption{{\it The case $g<0$}: The $(x,y)$-phase diagram of the
model, where $x= \mu_B|\vec B||g|$ and $y=eg^2B_{\perp}$, typical
for  values of $c\equiv e|g|/\mu_B>c^*\approx 28$. Physical region
of the diagram, corresponding to $B_\perp\le |\vec B|$ relation,
lies just below and/or to the right of the line L=$\{(x,y):~y=cx\}$.
Other notations are the same as in Fig. 7. }}
\end{figure}

\subsection{Numerical estimates in the context of condensed matter physics}

Now let us estimate the order of magnitude of the magnetic field at which the phase transitions of Figs 6, 7, 8 might take place in
(2+1)-dimensional condensed matter systems. To this end it is
necessary to take into account in the Lagrangian (1) the Fermi
velocity of quasi-particles $v_F\ne 1$ (see footnote 1 of the
present paper). Using the same calculational technique as in Sec. II
of the present paper and/or, e.g., in \cite{caldas,schakel,ramos},
it is possible to obtain the thermodynamic potential $\Omega_{v_F}$
for the case $v_F\ne 1$. Indeed, there is a very simple connection
between $\Omega_{v_F}$ and the renormalized TDP (\ref{2300})
corresponding to $v_F =1$. Namely, one should perform in
(\ref{2300}) the replacements $eB_\perp\to ev_F^2B_\perp$, $g\to
g/v_F$ (note, the Zeeman term $\mu_B|\vec B|$ remains unchanged in
this case) and then multiply the obtained expression by the factor
$1/v_F^2$.

Suppose that $g<0$ (recall, we still fix the spectroscopic Lande
factor $g_S$ by the relation $g_S=2$, as in graphene). Then, in the
particular case of $\vec B=0$ the TDP $\Omega_{v_F}$ thus obtained
from the TDP $V(M)$ (\ref{25}) of the case $v_F =1$ has already the
global minimum at the point $M_{0F}\equiv -v_F/g$ (it is the mass
gap of the system). Since in all numerical calculations of the case
$v_F =1$ an arbitrary dimensional quantity is converted into a dimensionless one by multiplying it with an appropriate powers of $|g|$,
in the case $v_F\ne 1$ the powers of $|g|/v_F$ should be used
instead. So, at $v_F\ne 1$ the analogs of the $(x,y)$-phase diagrams
of Figs 7, 8 are just the same figures, but with the new $x_F$-,
$y_F$-axes, where $x_F=x/v_F\equiv \mu_B|\vec B||g|/v_F$, and
$y_F=y$. (In the following, when referring to Figs 7, 8 in the case
$v_F\ne 1$, we imply that instead of $x$ and $y$ the new parameters
$x_F$ and $y_F$ should be used in these figures.) The line L, below
which the physical region is arranged, has the form $y_F=c_Fx_F$,
where $c_F\equiv cv_F=e|g|v_F/\mu_B=ev_F^2/(\mu_BM_{0F})$. It is
clear from Figs 7, 8 that at $B_\perp=0$ and $v_F\ne 1$ the phase
transition of the first order occurs at in-plane magnetic field
$|\vec B_0|$ corresponding to $x_F=1$, i.e. $|\vec
B_0|=v_F/(|g|\mu_B)=M_{0F}/\mu_B$. Since the value of the mass gap
$M_{0F}$ in condensed matter systems is typically of the order of
1-10 meV, one can easily obtain that the magnitude of the critical
magnetic field $|\vec B_0|$ is of order of 14-140 Teslas,
correspondingly. \footnote{In our numerical estimates we use the
following relations (see, e.g., in \cite{ramos}): $\mu_B=e/(2m_e)$,
where $m_e$ is the electron rest mass, $m_e\approx 0.5$ MeV; 1 Tesla
$\approx$ 700 eV$^2$; $e\approx 1/\sqrt{137}$, as in graphene. } It
is clear from Figs 7, 8 that at $B_\perp\ne 0$ the magnitudes of
$|\vec B|$, at which one can observe phase transitions, are even
less and might be as small as 0.7$|\vec B_0|$.

If $v_F=1/300$ and $g_S=2$, as in graphene, then the slope factor
$c_F$ of the line L is approximately equal to 10$^3$ at $M_{0F}=10$
meV, whereas it is of order of 10$^4$ at $M_{0F}=1$ meV, i.e.
$c_F\gg c^*\approx 28$. Hence, just the phase diagram of Fig. 8
refers to graphene-like planar systems. In this case (recall, we
assume that the critical line $l$ of the figure can be extrapolated
to the region of high $y$-values by a straight line with the slope
coefficient $c^*\approx 28$) chiral symmetry cannot be restored by
an arbitrary strong external perpendicular magnetic field, and the
enhancement effect is realized at $B_\perp\lessapprox |\vec B|$.
However, tilting the magnetic field away from a normal of the plane,
it is possible to restore the symmetry, if $|\vec B|>0.7|\vec B_0|$.
The angle $\varphi_0$ between $\vec B$ and the plane of the system,
at which the restoration of the symmetry occurs, can be estimated
numerically. For example, at $|\vec B|=1.5|\vec B_0|$ (i.e. at
$x_F=1.5$) the $B_{\perp}$-component, at which the chiral symmetry
is restored, corresponds to the value $y\approx 30$ (the point
$(1.5,30)$ lies on the critical curve $l$ of Fig. 8). Hence,
$\sin\varphi_0 =B_{\perp}/|\vec B|\approx 30/(1.5c_F)$. It means
that at $M_{0F}=10$ meV and $M_{0F}=1$ meV we have
$\sin\varphi_0\approx 0.02$ and $\sin\varphi_0\approx 0.002$,
correspondingly, i.e. the restoration of the chiral symmetry occurs
at very weak $B_{\perp}$-components of the magnetic field.

It was noted in the paper \cite{schakel}, where the properties of
the planar system of free electrons were investigated, that in the
gapless semiconductors Fermi velocity $v_F$ may vary in the interval
$v_F\in (1/3000,1/300)$, whereas the spectroscopic Lande factor $g_S$ might be as large as 200. The spectroscopic Lande factor $g_S\ne 2$ can be introduced in our consideration simply by re-scaling the Bohr
magneton, $\mu_B\to g_S\mu_B/2$. So, if $g_S=200$ and $v_F=1/300$,
as in \cite{schakel}, then both the modulus of the critical magnetic
field $|\vec B_0|$ and the slope factor $c_F$ of the line L take the
values which are 100 times smaller than in the case $g_S=2$, i.e. at
$M_{0F}=10$ meV we will have $|\vec B_0|=1.4$ Teslas and
$c_F=10<c^*\approx 28$. It means that in this case the phase
portrait of the model looks like in Fig. 7. As a result, the chiral
symmetry is restored at external (even perpendicular) magnetic
fields such that $|\vec B|\sim |\vec B_0|=1.4$ Teslas. Moreover, in
this case it is possible to reach the paramagnetic chiral symmetry
breaking phase 3 (see Figs 7 and 8) at not too small tilting angles.

If in addition to $g_S=200$ and $v_F=1/300$ we consider a system
with $M_{0F}=1$ meV, then $|\vec B_0|=0.14$ Teslas,
$c_F=100>c^*\approx 28$ and hence, as in graphene-like systems, the
relevant phase diagram is one of Fig. 8. However, in this case the
tilting angle $\varphi_0$ at which chiral symmetry can be restored
is 100 times larger. For example, at $|\vec B|=1.5|\vec B_0|$ the
angle $\varphi_0$, at which the restoration occurs, is defined by
the relation $\sin\varphi_0\approx 0.2$, i.e. $\varphi_0\approx
10^o$.

Note, up to now we have estimated phase transitions in the systems
with $v_F=1/300$. However, still smaller values of the critical
magnetic field $|\vec B_0|$ are realized in the planar gapless
semiconductors at smaller values of $v_F$, e.g., at $v_F=1/3000$. In
addition, in this case the slope factor $c_F$ of the line L might be
extremely small, i.e. $c_F\sim 1$. So, just the phase diagram of Fig.
6 with a variety of phase transitions is relevant for such condensed
matter systems.

In conclusion, we see that the effects which are due to the Zeeman
interaction can be observed in real condensed matter systems at
reasonable laboratory magnitudes of external magnetic fields.

\section{Summary and conclusions}

In the present paper we investigate (at zero temperature and
chemical potential) the response of the (2+1)-dimensional GN model
(1) upon the action of external magnetic field $\vec B$. The model
describes a four-fermion self-interaction of quasi-particles
(electrons) with spin 1/2. In addition, it describes the interaction of $\vec B$ both with orbital angular momentum of electrons and with their spin. The last is known as the Zeeman interaction, and it is proportional to electron magnetic moment $\mu_B$ which is a free model parameter in our consideration. So at $\mu_B=0$ the properties of the model were considered, e.g., in \cite{klimenko,Semenoff:1998bk,zkke},
where in particular it was established that an external
perpendicular magnetic field $\vec B_\perp$ induces spontaneous
chiral symmetry breaking at $G<G_c$, or it enhances chiral
condensation at $G>G_c$. (Such an ability of an external magnetic
field is called the magnetic catalysis effect.) Moreover, in this
case the system responds diamagnetically on the influence of
external magnetic field, i.e. its magnetization is negative. In
addition, there are no magnetic oscillations of any physical
quantity if the Zeeman interaction is not taken into account.

In the paper we study the modifications that appear both in the
magnetic catalysis  effect and in the magnetization phenomena of the
system when Zeeman interaction is taken into consideration, i.e. at
$\mu_B\ne 0$. To this end, we have obtained in the leading order of
the large-$N$ expansion technique the renormalized thermodynamic
potential $\Omega^{ren} (M;\nu,B_\perp)$ (\ref{2300}), where
$\nu=\mu_B|\vec B|$. The behavior of the global minimum point of
this quantity with respect to $M$ defines the phase structure of the
model, whereas its derivative with respect to $|\vec B|$ gives us
the magnetization. Note also that the renormalized TDP (\ref{2300})
depends no more on the bare coupling $G$. Instead, it appears the dependence of the TDP on
the new finite parameter $g$, which is connected with $G$ by the
relation (\ref{18}). (Note, it follows from (\ref{18})
that the values $g>0$ ($g<0$) correspond to the region $G<G_c$
($G>G_c$).) The main results of our investigations are the
following.

i) We have found that at $\mu_B\ne 0$ and $g>0$ there is a critical
coupling constant  $g_c=2\mu_B/e$ such that at $g>g_c$ an arbitrary
rather weak external magnetic field $\vec B$ induces spontaneous
chiral symmetry breaking provided that there is not too great a
deviation of $\vec B$ from a vertical as well as that $|\vec
B|<B_c(g)$, where $0<B_c(g)<\infty$ (see Fig. 2). At $0<g<g_c$
chiral symmetry cannot be broken by an external magnetic field.  (In
contrast, at $\mu_B=0$ and any values of $g>0$ the chiral symmetry
breaking is induced by arbitrary external magnetic field  $\vec B$
such that $\vec B_\perp\ne 0$.)

ii) Suppose that $\mu_B\ne 0$, $g>g_c>0$ and chiral symmetry is
broken, i.e. $\vec B$  has a rather large $B_\perp$ component. Then
chiral symmetry can be restored simply by tilting magnetic field to
a system plane, i.e. without any increase of its modulus $|\vec B|$.

iii) We have shown that at $\mu_B\ne 0$, $g>0$ and arbitrary fixed
$|\vec B|\ne 0$ one  can observe oscillations of the magnetization
in the region of small values of $B_\perp$ (see Figs 3 and 4). Note,
de Haas -- van Alphen magnetic oscillation phenomenon is rather
typical for condensed matter physics \cite{osc,osc2} as well as for
dense relativistic matter \cite{elmfors,vshivtsev,sadooghi}.
It occurs usually at nonzero chemical potential. In contrast, in
our (2+1)-dimensional system (1) this phenomenon is induced (at zero
chemical potential) by tilting an external magnetic field only.

iv) If $\mu_B\ne 0$ and $g<0$, then the phase structure and magnetic
properties of the model are much richer than in the case of $\mu_B=
0$,  $g<0$. Indeed, it is clear from Figs 6, 7 and 8 that at
non-vanishing Zeeman interaction the phase portrait of the model
contains at least two chirally nonsymmetric phases, denoted as 2 and
3. In the phase 2, which is a diamagnetic one, the enhancement of the
chiral symmetry is occurred, whereas in the paramagnetic phase 3 it
is absent. Moreover, if in addition the parameter $c\equiv
e|g|/\mu_B<c^*\approx 28$, then at sufficiently high values of
$|\vec B|$ (even at a perpendicular magnetic field) the restoration
of the chiral symmetry is occurred in the model. In contrast, at
$\mu_B= 0$ and $g<0$ only the diamagnetic phase 2 with enhancement of
the chiral symmetry breaking is realized in the model at arbitrary
values and directions of $\vec B$, such that $B_\perp>0$.

v) Assuming that the critical line $l$ of Fig. 8 can be extrapolated
to the region $y\equiv eg^2B_\perp>11$ by a straight line with a
slope coefficient $c^*\approx 28$, we see that at $g<0$ and $c\equiv
e|g|/\mu_B>c^*$ the line L of Fig. 8 does not cross any of the
critical curves of the figure. So, in this case at an arbitrary
perpendicular magnetic field chiral symmetry cannot be restored.
However, tilting the magnetic field away from a normal position, it
is possible to restore the symmetry. As our numerical estimates show
(see in Sec. IV C), just this situation is typical for graphene-like
planar systems. \footnote{If our assumption is not valid and the
line L of Fig. 8 crosses nevertheless the critical curve $l$ in the case of $e|g|/\mu_B>c^*$, then this can happen at extremely high values of $|\vec B|$,
which are not achievable in laboratories at present time. For
example, we see on the basis of numerical estimates of Sec. IV C
that in the graphene-like systems, for which the slope factor
$e|g|/\mu_B$ of the line L is very large, the restoration of the
chiral symmetry in this case might be realized by perpendicular
magnetic fields such that $|\vec B|\gg 140$ Teslas. }

vi) Look again at Fig. 8, where $c>c^*$, and fix $|\vec B|$, e.g., in the interval such that $0.8<x\equiv\mu_B|\vec B||g|<1$. Then at $B_\perp\lessapprox |\vec B|$ the chirally broken phase 2 is realized in the model. It is clear from the figure that decreasing the value of $B_\perp$ (or simply tilting $\vec B$), it is possible in this case to restore the chiral symmetry. However, a further reduction of $B_\perp$ leads to a transition of the system to the chirally broken phases 3 and 2. In contrast, if $|\vec B|$ is fixed in the interval corresponding to $1<x<1.3$, then after a chiral symmetry restoration, taking place for certain intermediate values of $B_\perp$, there should appear the chiral symmetry broken phase 3 and then, finally, again the symmetric phase of the model. Hence, at some fixed values of $|\vec B|$ a series of phase transitions, following one after another, can be caused in the system simply by changing the inclination of an external magnetic field.

While our work  was in the preparation, we learned about the paper
\cite{ramos}, where the same model with a tilted  magnetic field has
been investigated. However, in \cite{ramos} quite other properties
of  the model, such as quantum Hall effect etc, were studied at
nonzero temperature and chemical potential.

\end{document}